\documentclass{article}

\usepackage[english]{babel}

\usepackage[a4paper,top=2cm,bottom=2cm,left=3cm,right=3cm,marginparwidth=1.75cm]{geometry}

\usepackage{amssymb,amsmath,color,graphicx,dsfont,mathrsfs,amsthm,stmaryrd,mathtools}
\usepackage{authblk}
\usepackage{xcolor}
\usepackage{enumitem}
\usepackage[unicode,colorlinks=true,linkcolor=red,citecolor=blue,linktocpage=true]{hyperref}
\usepackage{braket}
\usepackage[capitalise]{cleveref}
\let\C\relax 

\usepackage{thmtools}
\usepackage{thm-restate}

\newcommand\rememberEqLabel[2]
  {\expandafter\gdef\csname labeled:#1\endcsname{#2}%
   \label{#1}#2}
\newcommand\recallEq[1]
   {\csname labeled:#1\endcsname\tag{\ref{#1}}}

\newtheorem{theorem}{Theorem}
\newtheorem{lemma}[theorem]{Lemma}

\newtheorem{definition}[theorem]{Definition}
\newtheorem{proposition}[theorem]{Proposition}

\makeatletter
\if@cref@capitalise
\crefname{hypothesis}{Hypothesis}{Hypotheses}
\else
\crefname{hypothesis}{hypothesis}{hypotheses}
\fi
\makeatother
\Crefname{hypothesis}{Hypothesis}{Hypotheses}

\usepackage[
    backend=biber,
    maxbibnames=6,
    maxcitenames=6,
    style=alphabetic,
    sorting=nyt,
    isbn=false,
    doi=true,
    url=false,
       firstinits=true,
    date=year]{biblatex}
\newcommand{\holofun}{\mathfrak H}

\newcommand{\N}{\mathbb{N}}

\newcommand{\C}{\mathbb{C}}
\newcommand{\R}{\mathbb{R}}

\renewcommand{\H}{\mathcal{H}}
\newcommand{\E}{\mathcal{E}}
\newcommand{\D}{\mathcal{D}}
\newcommand{\K}{\mathcal{K}}
\newcommand{\T}{\mathcal{T}}
\renewcommand{\S}{\mathcal{S}}
\renewcommand{\L}{\mathcal{L}}
\newcommand{\F}{\mathcal F}

\newcommand{\eqF}{=_\F}

\newcommand{\Id}{\mathds{1}}
\newcommand{\Tr}[1]{\operatorname{Tr}\left( #1 \right)}
\newcommand{\Span}[1]{\operatorname{Span}\left\{ #1 \right\} }
\renewcommand{\ker}[1]{\operatorname{Ker}\left( #1 \right)}

\renewcommand{\Re}{\operatorname{Re}}

\newcommand{\kb}[1]{\Ket{#1}\Bra{#1}}

\newcommand{\norm}[1]{\| #1 \|}
\newcommand{\trnorm}[1]{\| #1 \|_1}
\newcommand{\hsnorm}[1]{\| #1 \|_2}

\let\bf\boldsymbol

\newcommand{\orho}{{\bf \rho}}
\newcommand{\oa}{{\bf a}}
\newcommand{\ob}{{\bf b}}
\newcommand{\oor}{{\bf r}} 
\newcommand{\oB}{{\bf B}}
\newcommand{\oD}{{\bf D}}
\newcommand{\oL}{{\bf L}}
\newcommand{\oH}{{\bf H}}
\newcommand{\oG}{{\bf G}}
\newcommand{\oPi}{{\bf \Pi}}
\newcommand{\oP}{{\bf P}}
\newcommand{\oS}{{\bf S}}
\newcommand{\oC}{{\bf C}}
\newcommand{\oX}{{\bf X}}
\newcommand{\oY}{{\bf Y}}
\newcommand{\oA}{{\bf A}}
\newcommand{\oU}{{\bf U}}
\newcommand{\oPhi}{{\bf \Phi}}
\newcommand{\oV}{{\bf V}}
\newcommand{\oW}{{\bf W}}
\newcommand{\oR}{{\bf R}}
\newcommand{\oO}{{\bf O}}
\newcommand{\ox}{{\bf x}}

\newcommand{\proj}{\oPi}
\newcommand{\projE}{\oPi_E}
\newcommand{\projL}{\oPi_\oL}

\newlist{hypenum}{enumerate}{2} 
\setlist[hypenum]{label=(H\arabic*), ref=(H\arabic*), leftmargin=4em}
\crefalias{hypenumi}{hypothesis}

\binoppenalty=\maxdimen
\relpenalty=\maxdimen

\title{%
		Convergence of bipartite open quantum systems stabilized by reservoir engineering\\
}
\author[1]{Rémi Robin\thanks{remi.robin@minesparis.psl.eu}}
\author[1]{Pierre Rouchon\thanks{pierre.rouchon@minesparis.psl.eu}}
\author[1]{Lev-Arcady Sellem\thanks{lev-arcady.sellem@minesparis.psl.eu}}
\affil[1]{Laboratoire de Physique de l’\'Ecole Normale Supérieure, Mines Paris, Inria, CNRS, ENS-PSL, Sorbonne Université, PSL Research University, Paris, France}

\begin{document}
\maketitle

\abstract{%
We study a generic family of Lindblad master equations modeling bipartite open quantum systems,
where one tries to stabilize a quantum system by carefully designing its
interaction with another, dissipative, quantum system---a strategy
known as \emph{quantum reservoir engineering}.
We provide sufficient conditions for convergence of the considered Lindblad equations;
our setting accommodates the case where steady-states are not unique
but rather supported on a given subspace of the underlying Hilbert space.
We apply our result to a Lindblad master equation
modeling engineered multi-photon emission and absorption processes,
a setting that received considerable attention in recent years
due to its potential applications
for the stabilization of so-called \emph{cat qubits}.
}

\tableofcontents

\clearpage

\section{Introduction}
Closed quantum systems, isolated from their environment and governed by Schrödinger's
equation, follow a unitary evolution.
This unitarity prevents the engineering of stabilizing dynamics due to the preservation of distances.
For control purposes,
it is thus often interesting to rather use open quantum system not fully isolated from their environment,
where the coupling to the environment
provides the dissipation required for stabilization.
This idea, known as \emph{dissipation engineering} or \emph{quantum reservoir engineering}
\cite{%
poyatosQuantumReservoirEngineering1996,%
carvalhoDecoherencePointer2001,%
},
can be traced back to optical pumping and coherent population trapping
\cite{%
kastlerpumping,%
arimondoNonabsorbingAtomic1976,%
arimondoCoherentPopulation1996,%
}.
We will consider open quantum systems governed by a
Lindblad-type master equation,
also known as the Gorini–Kossakowski–Sudarshan–Lindblad
master equation
\cite{GKS_paper,lindbladGeneratorsQuantumDynamical1976,gkls_history},
which is notably suited to the study of systems weakly coupled to
a Markovian environment \cite{breuerTheoryOpen2006}.
In this formalism, the state of the system is given by a
density operator $\orho$,
that is a
self-adjoint, positive and trace-class operator
on a Hilbert space $\H$ with $\Tr{\orho}=1$,
satisfying
\begin{equation}
\label{eq-lindblad_general}
\frac d{dt} \orho_t
	= -i \, [ \oH, \orho_t]
		+ \sum_{\nu} D[\oL_\nu](\orho_t).
\end{equation}
Here,
$\oH$ is a self-adjoint unbounded operator called the Hamiltonian,
$[\oX,\oY] = \oX\oY-\oY\oX$ denotes the commutator of two operators,
the $\oL_\nu$ are unbounded operators called the loss operators or Lindblad operators,
and $D[\oL](\oX) = \oL \oX \oL^\dag - \tfrac12 \oL^\dag \oL \oX - \tfrac12 \oX \oL^\dag \oL$
where $\oL^\dag$ is the adjoint of $\oL$
(implicitly, we only consider operators that are densely defined and closable,
and we work with the convention $\hbar=1$).

For reservoir engineering purposes,
a fundamental question is to determine what states can be stabilized
assuming that part of the Hamiltonian and Lindblad operators
appearing in \cref{eq-lindblad_general} are dictated by the physics of the system,
whereas other parts can be tuned through external control inputs.
In particular, several proposals for the protection of quantum information
using reservoir engineering techniques
rely on the stabilization of so-called \emph{decoherence-free subspaces}:
in that setting, the engineered dissipation cancels on a multidimensional subspace
$\H_0\subset \H$,
and the set of density operators $\orho$ supported on $\H_0$,
that is satisfying $\Pi_0 \orho \Pi_0 = \orho$ with $\Pi_0$ the orthogonal projection
onto $\H_0$,
is attractive.
In the simplest case, given a target subspace $\H_0\subset \H$,
the dynamics to engineer contains only one Lindblad operator
\begin{equation}
\label{eq-lindblad_onedissip}
\frac d{dt} \orho_t = D[\oL](\orho_t)
\end{equation}
chosen so that $\H_0 = \ker{\oL}$ on the one hand,
and solutions to \cref{eq-lindblad_onedissip} converge to density operators supported
on $\H_0$ on the other hand.

Note that while we mention the protection of quantum information,
reservoir engineering is also useful outside this realm
and has been studied, for instance, for active reset protocols
\cite{%
murchCavityAssistedQuantumBath2012,%
geerlingsDemonstratingDrivenReset2013,%
}
or generation of entangled states
\cite{%
pielawaGenerationEinsteinPodolskyRosenEntangledRadiation2007,%
krauterEntanglementGeneratedDissipation2011,%
shankarAutonomouslyStabilizedEntanglement2013,%
}.
In these cases, the subspace $\H_0$ is usually of dimension one.
This is covered by the analysis led in this paper,
but we chose to favor examples from the quantum computing literature where $\H_0$ can have
an arbitrary dimension,
leading to a richer problem.
\newline

We will consider in particular the example of multi-photon dissipative processes,
which provided the initial motivation for our study due to the renewed interest
it attracted in recent years for the stabilization of so-called \emph{cat qubits}
\cite{%
cochraneMacroscopicallyDistinct1999,%
leghtasHardwareefficientAutonomous2013,%
mirrahimiDynamicallyProtected2014,%
chamberlandBuildingFaultTolerant2022%
}.
The corresponding Hilbert space is $\H = L^2(\R,\C)$ and the subspace of interest
is
\begin{equation}
\H_0 = \Span{\ket{\omega^r \alpha} \; | \; \omega = e^{\frac{2i\pi}{k}}, 0\leq r \leq k-1}
\end{equation}
where $\alpha\in \mathbb C$ and $k\in\N^*$ are parameters
and for any complex number $z\in\C$,
the so-called \emph{coherent state} $\ket z$
is defined as the normalized eigenstate of the annihilation operator
$\oa = (x + \partial_x)/\sqrt 2$
associated with the eigenvalue $z$;
in the usual Fock basis $(\ket n)_{n\in\N}$ of $L^2(\R,\C)$ it takes the form
\begin{equation}
\ket z = e^{-|z|^2/2} \sum_{n=0} \frac{z^n}{\sqrt{n!}}\ket n.
\end{equation}
Choosing $\oL = \oa^k-\alpha^k \Id$,
one easily checks that $\ker\oL = \H_0$.
Formal investigations of the convergence of
\cref{eq-lindblad_onedissip}
for this choice of operator $\oL$
date back to \cite{gerryGenerationEven1993}.
More recently, a rigorous study of its
well-posedness and convergence properties was conducted in
\cite{azouitWellposednessConvergenceLindblad2016};
in particular, the convergence analysis relying on Lyapunov techniques provides explicit
exponential convergence rates toward the set of density operators supported on $\H_0$.
\newline

In view of experimental implementations,
one faces the difficulty that exotic dissipation processes canceling on a prescribed
subspace $\H_0$ (such as the multi-photon dissipation examples above)
are usually not readily available as natural physical processes.
More often, dissipative processes are imposed by the specific physical platform under study
(such as atomic decay for atomic systems, light emission and absorption for
electromagnetic systems, etc.)
while the Hamiltonian part can contain control degrees of freedom modeling
engineered interactions between subsystems.
The engineering of exotic dissipative processes thus has to be approximated as
an exotic coupling to an ancillary system, which is itself subject to natural dissipative processes.
This strategy motivates a particular emphasis on the study of \emph{bipartite}
open quantum systems,
composed of two coupled subsystems where one subsystem is the actual system of interest
whereas the other is an ancillary system used to circumvent the impossibility of directly
engineering dissipative processes.
For instance, for the generation and stabilization of cat qubits,
experimental proposals and demonstrations
resort to repeated interactions with a stream of two-level ancillae
or an actively reset ancilla
\cite{%
sarletteStabilizationNonclassicalStates2011,%
ofekExtendingLifetime2016,%
},
or to nonlinear couplings with a strongly damped electromagnetic mode
\cite{%
leghtasConfiningStateLight2015,%
touzardCoherentOscillations2018,%
lescanneExponentialSuppression2020,%
berdouOneHundred2023,%
regladeQuantumControl2023,%
}.

The goal of the present paper is to give a sound theoretical grounding
to the latter solution.
Mathematically, we thus focus on a specific class of Lindblad master equations
modeling bipartite open quantum systems with dissipation on only one subsystem.
The corresponding Hilbert space has a tensor structure $\H = \H_a \otimes \H_b$;
here we will moreover only consider the case where $\H_a$ and $\H_b$
are two copies of $L^2(\R,\C)$,
and denote $\oa$ and $\ob$ the annihilation operators on $\H_a$ and $\H_b$.
The dynamics takes the form
\begin{equation}
	\label{eq-intro_maineq}
	\frac d{dt} \orho_t = -i \, [\oL \otimes \ob^\dag + \oL^\dag \otimes \ob, \orho_t ]
				+ \kappa \, D[\Id_a \otimes \ob](\orho_t)
\end{equation}
where $\kappa>0$ is a parameter and $\oL$ is, at this stage, an arbitrary operator
on $\H_a$.
In terms of physical interpretation,
the only loss operator, proportional to the
annihilation operator on subsystem $b$,
can model for instance
photon losses in an electromagnetic mode
(usually the dominant dissipative process in this setting).
A justification for the model in \cref{eq-intro_maineq}
is found in the theory of adiabatic elimination:
in the limit of strong dissipation $\kappa\gg 1$,
the solution to \cref{eq-intro_maineq} is pertubatively approximated as a series in $1/\kappa$.
A formal calculation
(see for instance \cite{azouitAdiabaticElimination2017})
shows that, to second order,
the solution of \cref{eq-intro_maineq} can be approximated as
\begin{equation}
	\label{eq-adiab_link}
	\left\{
		\begin{aligned}
			\orho_t &\simeq \K \big(\orho^a_t \otimes \ket0\bra0\big),\\
			\frac d{dt} \orho^a_t &= \tilde \kappa D[\oL](\orho^a_t)
		\end{aligned}
	\right.
\end{equation}
where $\K$ is a Kraus map close to identity, $\tilde \kappa = 4/\kappa$
and $\orho^a$ is a density operator on $\H_a$ only.
In other words, when the dissipation on the ancillary system $b$ is strong enough,
the solution to \cref{eq-intro_maineq} stays close to a separable density operator
$\orho_t = \orho^a_t \otimes \orho^b_t$ (with $\orho^b_t = \ket0\bra0$),
describing a system where the ancillary subsystem $b$ stays in vacuum
whereas the subsystem $a$ follows \cref{eq-lindblad_onedissip} with the engineered
dissipation operator $\oL$.
\newline

In light of the above, a natural question is then:
given a subspace $\H_0\subset \H_a$
and an
operator $\oL$ on $\H_a$ such that $\ker\oL = \H_0$,
what conditions ensure that
every solution $\orho$ to \cref{eq-intro_maineq} converge to a separable density operator
$\orho_\infty = \orho^a_\infty \otimes \orho^b_\infty$
such that $\orho^b_\infty = \ket0\bra0$ and $\orho^a_\infty$ is supported on $\H_0$?

Considering adiabatic elimination,
a tempting initial guess could be to require that $\oL$
be chosen so that every solution $\orho^a$ to \cref{eq-lindblad_onedissip}
converges to a density operator $\orho^a_\infty$ supported on $\H_0$.
However, we emphasize that we are interested here in the properties of \cref{eq-intro_maineq}
in and of itself. We dismiss the fact that it can be linked with \cref{eq-lindblad_onedissip},
and in particular our analysis does \emph{not} use the adiabatic elimination nor require
$\kappa \gg 1$.
We propose three arguments supporting the study of \cref{eq-intro_maineq}
independently of the approximation that initially motivated its introduction:
\begin{enumerate}
	\item Formally, using $\ker\oL = \H_0$,
		any operator $\orho^a \otimes \ket0\bra0$ with $\orho^a$ supported on $\H_0$
		is a steady-state of \cref{eq-intro_maineq}.
		We are thus in a very peculiar situation where the exact trajectories could be
		approximated by adiabatic elimination within an accuracy depending on $\kappa$,
		but the steady-states are actually independent of $\kappa$.
		It is thus reasonable to hope that a convergence result can be obtained
		without assuming $\kappa\gg 1$;
	\item As explained above, in physical implementations, direct dissipation engineering is
		impossible so that \cref{eq-intro_maineq} is a model closer than
		\cref{eq-lindblad_onedissip} to existing experiments,
		and its study can help understand these
		experiments better;
	\item We see in \cref{eq-adiab_link} that in the regime $\kappa\gg 1$ where
		the adiabatic elimination is valid,
		the engineered dissipation strength $\tilde \kappa = 4/\kappa$ is small.
		It is thus tempting to explore the properties of \cref{eq-intro_maineq} outside
		this regime.
\end{enumerate}

A consequent mathematical literature has been devoted to the study of GKSL equations,
using the formalism of Quantum Dynamical Semigroups
\cite{%
daviesQuantumDynamicalSemigroups1977a,%
}
or Quantum Markov Semigroups
\cite{%
fagnolaQuantumMarkovSemigroups1999,%
}.
However, to the best of our knowledge,
studies of generic convergence conditions
generally either consider
finite-dimensional systems
\cite{%
baumgartnerAnalysisQuantum2008a,%
baumgartnerAnalysisQuantum2008b,%
},
or focus on existence of invariant states
and in particular on systems with a unique and/or faithful invariant state
\cite{%
fagnolaExistenceStationaryStates2001,%
fagnolaQuantumMarkovSemigroups2003,%
}.
In order to cover relevant examples from the physics literature,
including the multi-photon dissipative processes presented above in connection with the stabilization
of cat qubits,
we need to obtain convergence results in infinite dimension,
for GKSL equations with unbounded generators,
and featuring multiple steady-states supported on a given subspace of the underlying Hilbert space
and thus not faithful (usually the subspace has finite dimension, in which case the invariant states
we will consider are even finite rank).

Our study of \cref{eq-intro_maineq} is organized as follows.
\cref{sec-preliminaries} covers technical preliminaries:
\cref{sec-intro_notations} gathers notations
and
\cref{section-definition-lindblad}
recalls the definition of solution to a Lindblad master equation
and presents a few existence theorems from the literature.
\cref{sec-lasalle_mainsec} contains the main result of this paper
in the form of \cref{th-main_convergence},
where we provide a set of sufficient conditions to ensure
that the solutions of \cref{eq-intro_maineq}
converge to density operators supported on $\ker\oL \otimes \ket 0$.
These conditions are briefly discussed in \cref{ssec-hypdisc}:
broadly speaking, they allow establishing convergence from
the existence of a suitable energy operator bounded along trajectories
together with an algebraic condition of density of a suitable subspace
in $\H$.
In \cref{ssec-thm1proofsketch}, we sketch the proof and dice it into a few  main steps,
before providing the full proof of each step in \cref{ssec-proof_mainthm}.
\cref{sec-catqubits} presents the application of \cref{th-main_convergence}
to the study of multi-photon dissipative processes, that is for the choice $\oL = \oa^k -\alpha^k \Id$
where $\alpha\in\C$ and $k\in\N^*$ are parameters.
This study shows that the sufficient conditions we propose can be checked on a physically relevant
example;
to the best of our knowledge, it is the first proof of convergence for this model.
Among the conditions to check, the density condition proves harder to establish.
Interestingly, it is linked to complex analysis considerations,
more precisely to a problem of Newman and Shapiro about polynomial approximation
in the Bargmann-Fock space of holomorphic functions.
For pedagogy purposes, we first present the case where $\alpha=0$ or $k=1$, where no such complication
occurs, before solving the generic case.
Finally, \cref{sec-conclusion} presents our conclusions and perspectives for future work.

\section{Technical preliminaries}
\label{sec-preliminaries}
\subsection{Notations}
\label{sec-intro_notations}
We fix $\hbar=1$ and work with dimensionless quantities. Besides, we use the following notations
\begin{itemize}
    \item $\H$ is a complex separable Hilbert space. Scalar products are denoted using Dirac's bra-ket notation, namely $\ket{x}$ is an element of $\H$, whereas $\bra{x}$ is the linear form canonically associated to the vector $\ket{x}$.
	    \item Operators on $\H$ are denoted with bold characters such as $\oa, \ob, \orho$.
    \item $\K^1$ or $\K^1(\H)$ is the Banach space of trace-class operator on $\H$,
	    equipped with the trace norm defined by:
    \begin{align*}
	    \forall \orho \in \K^1, \quad \trnorm\orho=\Tr{\sqrt{\orho^\dag \orho}}.
    \end{align*}
    \item $\K^2$ or $\K^2(\H)$ denotes the space of Hilbert–Schmidt operators on $\H$,
	    equipped with the Hilbert-Schmidt norm defined by:
    \begin{align*}
	    \|\orho\|_2=\sqrt{\Tr{\orho^\dag \orho}}.
    \end{align*}
	\item $\K_d\subset \K^1$ denotes the convex set of density operators, i.e.,
    \begin{align*}
        \K_d= \left\{ \orho \in \K^1 \mid \orho^\dag= \orho,\, \Tr{\orho}=1, \, \orho\geq 0 \right\}.
    \end{align*}
    \item $B(\H)$ denote the (Von Neumann) algebra of bounded operators on $\H$.
	    $\Id$ or $\Id_\H$ denotes the identity operator.
    \item If $\oA$ is a (unbounded) linear operator on $\H$, we denote by $\D(\oA)$ its domain,
	    and define $\D(\oA^\infty) = \cap_{n\geq0} \D(\oA^n)$. When the Hilbert space has to be specified, we denote the operator by the triplet $(\oA,\D(\oA),\H)$.
    \item When working on the Hilbert space $\H = L^2(\R,\C)$, we define
	    the so-called annihilation operator
	    $\oa = \tfrac1{\sqrt2} (x+\partial_x)$.
		Similarly, when working with two copies $\H_a$ and $\H_b$ of $L^2(\R,\C)$,
		we will denote respectively $\oa$ and $\ob$
		the annihilation operators relative to $\H_a$ and $\H_b$.
	\item When working on a tensor product space $\H= \H_a \otimes \H_b$,
		given two operators $\oX_a$ and $\oX_b$
		defined respectively in $\H_a$ and $\H_b$,
		we will often alleviate the notations by identifying them, respectively, to the operators
		$\oX_a\otimes \Id_{\H_b}$ and $\Id_{\H_a} \otimes \oX_b$ on $\H$.
		Similarly, we will write $\oX_a \oX_b$ for $\oX_a \otimes \oX_b$.

	\item For $n\geq 1$ and $A$ a polynomial ring, we denote
		$A\langle X_1, \ldots, X_n\rangle$ the free algebra on $n$ indeterminates over $A$;
		 equivalently,
		 it can be understood as the set of non-commutative polynomials
		 in $n$ indeterminates over $A$.
		In what follows, we will only use the case $A=\C$;
		when $n=2$, we will additionally use the notation $X,Y$
		instead of $X_1,X_2$ to denote the indeterminates.
\end{itemize}
\vspace{1em}

For the evolution of open quantum systems, we use the following conventions:
\begin{itemize}
    \item Let $\oL$ and $\oX$ be linear operators and $\orho\in \K^1$,
	    we define:
    \begin{align}
	    D[\oL](\orho) &= \oL \orho \oL^\dag
	    			-\frac{1}{2} \oL^\dag \oL \orho
				-\frac{1}{2}\orho \oL^\dag \oL\\
	    D^*[\oL](\oX) &= \oL^\dag \oX \oL
	    			-\frac{1}{2} \oL^\dag \oL \oX
				-\frac{1}{2} \oX \oL^\dag \oL,
    \end{align}
    \item $\mathcal{L}$ denotes the Lindbladian super-operator
	    associated to a Hamiltonian $\oH$
		and a family of so-called Lindblad operators
		$(\oL_j)_j$ on $\H$%
		\footnote{
			In the context of Lindblad equations,
			operators denoted with a capital $\oL$ usually denote Lindblad operators.
			This may first appear at odds with the generic Lindblad \cref{eq-intro_maineq}
			introduced earlier, where the operator $\oL$ contributes to the Hamiltonian part of the dynamics.
		This choice of notations is explained by the motivation previously proposed
		for the study of
		\cref{eq-intro_maineq}, drawing from its use in quantum reservoir engineering
		through adiabatic elimination.
		In this context, \cref{eq-intro_maineq} is introduced as an approximation
		of the physically unrealistic \cref{eq-adiab_link},
		where $\oL$ indeed appears as a Lindblad operator.
		},
		acting on elements $\orho$ in (a domain in) $\K^1$ through
    \begin{align}
        \mathcal{L}(\orho)= -i [\oH,\orho] + \sum_j D[\oL_j](\orho).
    \end{align}
    In this article, we are interested in the case where $\mathcal{L}$ is not bounded.
We refer to \cref{section-definition-lindblad} for a proper definition
		of the semigroup $(\S_t)_{t\geq0}$
		associated with $\mathcal L$,
		and denote $\orho_t=\S_t(\orho_0)$
		the solution of the dynamical system
		\begin{equation}
			\frac d{dt} \orho_t = \mathcal L(\orho_t)
		\end{equation}
		initialized in a given element $\orho_0\in\K^1$.

    \item $\mathcal{L}^*$ is formally the adjoint of $\mathcal{L}$;
	    for $\oX$ in (a domain in) $B(\H)$, it takes the form
    \begin{align}
        \label{eq-formal-lindblad_adjoint}
        \mathcal{L}^*(\oX)= i [\oH,\oX] +\sum_j D^*[\oL_j](\oX).
    \end{align}
    We refer again to \cref{section-definition-lindblad} for a proper definition
		of the associated semigroup $(\T_t)_{t\geq0}$,
		and denote $\oX_t = \T_t(\oX_0)$
		the solution of the adjoint dynamical system
		\begin{equation}
			\frac d{dt} \oX_t = \mathcal L^*(\oX_t)
			\label{eq:def_lindblad_heisenberg}
		\end{equation}
		initialized in a given element $\oX_0\in B(\H)$.
		Note that the semigroups
		$(\T_t)_{t\geq 0}$ on bounded operators
		and
		$(\S_t)_{t\geq 0}$ on trace-class operators
		are related through the following identity,
		where $t\geq 0$,
		$\oX\in B(\H)$ and $\orho_0\in \K^1$:
    \begin{align}
        \label{eq-semigroup-dual}
        \Tr{\S_t(\orho_0) \oX}=\Tr{\orho_0 \T_t(\oX)}.
    \end{align}
		Moreover, $(\S_t)_{t\geq 0}$ is called the pre-dual semigroup
		associated with $(\T_t)_{t\geq 0}$,
		since $B(\H)$ is the dual of $\K^1(\H)$
		(where the linear functional associated to $\oX \in B(\H)$
		is $\orho\mapsto \Tr{\oX\orho}$).

		In the physics literature, the evolution of density operators with $\S_t$
		is called the \emph{Schrödinger picture},
		while the evolution of operators with $\T_t$ is called the \emph{Heisenberg picture}.
\end{itemize}
\vspace{1em}

For holomorphic functions, we use the following notations:
\begin{itemize}
	\item $\holofun$ denotes the set of holomorphic functions on $\C$.
	\item $Z_f$ denotes the zero set of a holomorphic function $f$.
		For $k\geq 1$, $Z_f^k$ denotes the set of zeros of order $k$ of $f$.
	\item $\F^2$ denotes the Bargmann–Fock space,
		which is the set of holomorphic functions on $\C$ which belong
		to $L^2(\C, \tfrac1\pi e^{-|z|^2} dz)$
		with $dz$ the Lebesgue measure on the complex plane.
		It is a Hilbert space endowed with the inner product
		\(
			\braket{f|g}_{\F^2} =
				\frac1\pi \int_\C \overline{f(z)} \, g(z) \, e^{-|z|^2} dz.
		\)
\end{itemize}

\clearpage
\subsection{Well-posedness theorems}
\label{section-definition-lindblad}
One can choose between two equivalent definitions
of the solution to a Lindblad equation:
using the formulation on the Von Neumann algebra $B(\H)$
as done in Chebotarev and Fagnola's works
\cite{chebotarevSufficientConditionsConservativity1993,chebotarevSufficientConditionsConservativity1998},
or directly on the Banach space of trace class operators
as initially introduced by Davies
\cite{daviesGeneratorsDynamicalSemigroups1979,daviesQuantumDynamicalSemigroups1977a}.
In this paper, we take the first approach. Let us define the notion of Quantum Dynamical Semigroup (QDS):
\begin{definition}{}
    A quantum dynamical semigroup $(\T_t)_{t \geq 0}$ is a family of operators
	acting on $B(\H)$ which satisfies the following properties:
    \begin{itemize}
        \item $\T_0(\oX)=\oX$ for all $\oX\in B(\H)$,
        \item $\T_{t+s}(\oX)=\T_t(\T_s(\oX))$ for all $t,s\geq 0$ and $\oX \in B(\H)$,
        \item $\T_{t}(\Id)\leq \Id$ for all $t\geq 0$,
        \item $\T_{t}$ is a completely positive map for all $t\geq0$.
		This means that for any finite sequences
		    $(\oX_j)_{1\leq j \leq n}$
		    and
		    $(\oY_j)_{1\leq j \leq n}$
		    of element of $B(\H)$, we have
        \begin{align*}
            \sum_{1\leq j,l \leq n} \oY_l^\dag \, \T_t(\oX_l^\dag \oX_j) \,\oY_j \geq 0
        \end{align*}
\item (normality) for every weakly converging sequence $(\oX_n)_n \rightharpoonup X$
	in $B(\H)$,
		    the sequence $(\T_t(\oX_n))_n$
		    converges weakly toward $\T_t(\oX)$.
        \item (ultraweak continuity) for all $\orho \in \K^1$ and $\oX \in B(\H)$,
		we have
        \begin{align*}
            \lim_{t\to 0^+} \Tr{\orho \T_t(\oX)}=\Tr{\orho \oX}.
        \end{align*}
    \end{itemize}
\end{definition}

Let us now relate the Lindblad equation with the notion of quantum dynamical semigroup.
We introduce $\oG=-i\oH-\frac12\sum_{j} \oL_j^\dag \oL_j$
and assume that $\oG$ is the generator
of a strongly continuous semigroup of contractions for the Hilbert norm on $\H$,
that is of a semigroup $\Gamma\colon \R_+\mapsto \mathcal L(\H)$ satisfying:
\begin{itemize}
	\item $\Gamma(0) = \Id$,
	\item $\Gamma(t+s) = \Gamma(t)\, \Gamma(s) \quad \forall t,s\geq 0$,
	\item $\forall \ket u\in\H, \;
		\norm{\Gamma(t)\ket u - \ket u} \xrightarrow[t\rightarrow 0^+]{} 0$,
	\item $\norm{\Gamma(t)} \leq 1, \quad \forall t\geq 0.$
\end{itemize}
We say that the quantum dynamical semigroup $(\T_t)_{t\geq 0}$
is solution of \cref{eq:def_lindblad_heisenberg}
if and only if the following equation is satisfied:
\begin{align}
    \label{eq-Lindblad-precise}
    \Bra{v} \T_t(\oX) \Ket{u} = \Bra{v} \oX \Ket{u}
				+ \int_0^t \left( \bra{v} \T_s(\oX)  \ket{ \oG u}
				+ \bra{ \oG v} \T_s(\oX) \ket u
				+ \sum_j \bra{\oL_j v}  \, \T_s(\oX)\,   \ket{\oL_j u} \right) ds
\end{align}
for all $\ket u, \ket v \in \D(\oG)$,
$\oX\in B(\H)$ and $t\geq 0$.
Note that equivalently (\cite{chebotarevSufficientConditionsConservativity1998}[prop. 2.3]),
$(\T_t)_{t\geq 0}$ satisfies
\begin{align}
    \label{eq-lindblad-precise2}
    \Bra{v} \T_t(\oX) \Ket{u} &=
	\Bra{e^{t\oG} v} \oX \Ket{e^{t\oG} u}
		+\sum_j \int_0^t \Bra{\oL_j e^{(t-s) \oG} v}
			\, \T_s(\oX) \, \Ket{ \oL_j e^{(t-s)\oG}  u} ds,
\end{align}
for all $\ket u,\ket v \in \D(\oG)$, $\oX\in B(\H)$ and $t\geq 0$.

The existence of a quantum dynamical semigroup solution of \cref{eq-Lindblad-precise}
can be obtained for example with the following theorem:

\begin{theorem}{\cite[Theorem 3.22]{fagnolaQuantumMarkovSemigroups1999}}
    \label{lem-existence-semigroup}
	Assume that
		$\oG$ is the generator of a strongly continuous contraction semigroup,
		its domain satisfies $\D(\oG)\subset \cap_j \D(\oL_j)$,
		and for every $\ket u \in \D(\oG)$:
		\[
			\braket{ u |\oG u} + \braket{\oG u | u}
				+ \sum_j \braket{\oL_j u | \oL_j u} \leq 0.
		\]
		Then there exists a quantum dynamical semigroup $(\T_t^{min})_t$ solving \cref{eq-Lindblad-precise,eq-lindblad-precise2} such that
    \begin{itemize}
        \item $\T_t^{min}(\Id)\leq \Id$,
        \item for every $\sigma$-weakly continuous family $(\T_t)_t$ of positive map on $B(\H)$ solving \cref{eq-Lindblad-precise,eq-lindblad-precise2},
		and every $\oX\in B(\H)$, we have
        $$\T_t^{min}(\oX)\leq \T_t(\oX).$$
    \end{itemize}
\end{theorem}
The semigroup $(\T_t^{min})$ is called
the minimal semigroup
and one can easily prove
(\cite[Corollary 3.23]{fagnolaQuantumMarkovSemigroups1999})
that if the minimal semigroup $(\T_t^{min})_t$ satisfies
$\T_t(\Id)= \Id$ for all $t \geq 0$,
a property known as conservativity,
then it is the unique quantum dynamical semigroup solution of \cref{eq-Lindblad-precise}.
In that case, the semigroup is called a Quantum Markov Semigroup (QMS).

Necessary and sufficient conditions to ensure that the minimal semigroup of a Lindblad equation
is conservative
can be found in
\cite{chebotarevSufficientConditionsConservativity1993,%
chebotarevSufficientConditionsConservativity1998,%
fagnolaQuantumMarkovSemigroups1999}.
When it is the case, we will say that the Lindblad equation is well-posed.
In this paper, we will use in particular the following result to check conservativity of the minimal
semigroup:
\begin{theorem}{\cite[Theorem 3.40]{fagnolaQuantumMarkovSemigroups1999}}
    \label{thm-conservativity-semigroup}
	The minimal quantum dynamical semigroup is conservative under the following assumptions:
	\begin{enumerate}
		\item
		$\oG$ is the generator of a strongly continuous contraction semigroup,
		its domain satisfies $\D(\oG)\subset \cap_j \D(\oL_j)$,
		and for every $\ket u \in \D(\oG)$:
		\[
			\braket{u|\oG u}+ \braket{\oG u | u}+ \sum_j \braket{\oL_j u|\oL_j u} = 0.
		\]

		\item
		There exists a positive self-adjoint operator $\oPhi$ such that
			$\D(\oG) \subset \D(\oPhi^{1/2})$
			and for every $\ket u \in \D(\oG)$:
			\[ -2\Re(\braket{u| \oG u})
				= \sum_j \braket{\oL_j u|\oL_j u}
				= \left(\bra u \oPhi^{1/2}\right)\left(\oPhi^{1/2} \ket u\right).
			\]
		\item
		There exists a positive self-adjoint operator $\oC$ such that
			$\D(\oC) \subset \D(\oPhi)$
			and for every $\ket u\in \D(\oC)$:
			\[
				\left(\bra u \oPhi^{1/2}\right)\left(\oPhi^{1/2} \ket u\right)
				\leq \left(\bra u \oC^{1/2}\right)\left(\oC^{1/2} \ket u\right).
			\]

		\item
			There exists a linear manifold D which is a core for $\oC^{1/2}$,
			contained in $\D(\oG)$ and stable by the semigroup generated by $\oG$ and $\lambda_0 \geq0$,
			such that for any $\lambda>\lambda_0$ and $j\in\N$,
			\[
				\left(\oG -\lambda\Id\right)^{-1}(D) \subset \D(\oC^{1/2}),
				\quad
				\oL_j \left(\oG -\lambda\Id\right)^{-1}(D) \subset \D(\oC^{1/2});
			\]
		and there exists $\lambda>0$ and $b>0$ such that
			for every $\ket u \in \left(\oG -\lambda\Id\right)^{-1}(D)$:
			\[
				2\Re\left( \braket{\oC^{1/2}u|\oC^{1/2} \oG u}\right)
					+ \sum_j
						\braket{ \oC^{1/2} \oL_j u| \oC^{1/2} \oL_ju}
					\leq b \norm{\oC^{1/2}\ket u}^2
			\]
			Note that, apart from the technical assumptions on domains,
			this last hypothesis essentially boils down to:
			there exists $b>0$ and a positive self-adjoint operator $\oC$ such that
			$\mathcal L^*(\oC) \leq b \, \oC$.
	\end{enumerate}

\end{theorem}
To be precise,
note that the last assumption of this theorem
is stated with $\lambda_0=0$
in \cite[Theorem 3.40]{fagnolaQuantumMarkovSemigroups1999}.
However, their proof clearly indicates that only the domain inclusions involving the resolvent for sufficiently large $\lambda$ are needed, as stated in \cref{thm-conservativity-semigroup}.

Finally, a pre-dual semigroup $(\S_t)_{t\geq 0}$ on $\K^1$
is associated with $(\T_t)_{t\geq 0}$
through \cref{eq-semigroup-dual};
for a given initial condition $\orho_0\in\K^1$,
we will note $\orho_t = \S_t(\orho_0)$ the associated solution of the Lindblad equation.

\section{A LaSalle-like invariance principle}\label{sec-lasalle_mainsec}
As stated in the introduction,
we work on the Hilbert space
\( \H = \H_a\otimes\H_b \)
where $\H_a$ and $\H_b$ are two copies of $L^2(\R,\C)$,
and consider the following Lindblad equation
associated to a given (unbounded) operator $\oL$ on $\H_a$
and initial condition $\orho_0\in\K_d$:
\begin{align}
	\rememberEqLabel{eq-maineq}{
	\frac d{dt} \orho_t &= \mathcal L(\orho_t)
	= -i \left[\oH, \orho_t \right]
	+ \kappa \, D[\ob](\orho_t)}
\end{align}
where $\kappa>0$ is a parameter,
$\ob$ denotes the annihilation operator in $\H_b$
and $\oH = \oL \ob^\dag + \oL^\dag \ob$.
\newline

Denoting $\H_\oL = \ker\oL\otimes\ket0$
and $\projL$ the orthogonal projector onto $\H_\oL$,
our goal is to obtain a set of sufficient conditions
under which the solution $\orho_t$ of~\cref{eq-maineq}
converges to a density operator $\orho_\infty$ supported
on $\H_\oL$,
that is satisfying
$\projL \orho_\infty \projL = \orho_\infty$
or equivalently
$\rho_\infty \in \K^1(\H_\oL) \cap \K_d$.
The main intuition consists in exploiting the fact that if this convergence
holds,
then the quantity $\Tr{\projL \orho_t}$,
quantifying the probability mass supported in $\H_\oL$, should converge to $1$,
so that the function $t\mapsto 1-\Tr{\projL \orho_t}$
is a natural candidate Lyapunov function for the dynamics.
Making this basic idea rigorous allows us
to gather a set of sufficient conditions for convergence
in \cref{th-main_convergence},
which constitutes the main theorem of this article.
Before diving into the proof of the theorem,
we shortly discuss each hypothesis
in \cref{ssec-hypdisc}.
We then propose a rough sketch of the proof in \cref{ssec-thm1proofsketch}
followed by the detailed proof in \cref{ssec-proof_mainthm}.

\subsection{Main theorem}
\label{ssec-mainthm}

\begin{theorem}
    \label{th-main_convergence}
	Denoting $\oG = -i\oH - \frac\kappa2\ob^\dag\ob$,
	assume the following hypotheses hold:
	\begin{hypenum}
	\item \label{hyp-wellposed}
		Equation (\ref{eq-maineq}) is well-posed,

	\item \label{hyp-domain}
		There exists a subspace $D$ with
		$\H_\oL \subset D \subset \D\left( (\oG^\dag)^\infty\right)$
		invariant by both $\ob^\dag$ and the semigroup generated by $\oG^\dag$,

	\item \label{hyp-tightness}
		For any $\orho_0\in \K_d$ and $\epsilon>0$, there exists a linear subspace $E$
	satisfying:
		\begin{equation*}
			\left\{
				\begin{aligned}
								& \dim(E) < \infty,\\
				& \forall t \geq 0, \; \Tr{\projE  \orho_t}>1-\epsilon,
				\end{aligned}\right.
		\end{equation*}
		where $\orho_t$ is the solution of \cref{eq-maineq}
		starting from $\orho_0$
		and
		$\projE$ is the orthogonal projector on $E$,

	\item \label{hyp-density}
		The subspace
		\begin{equation}
		\E_\oL = \Span{ P(\oG^\dag,\ob^\dag) \Ket{v}\otimes \Ket{0}
			\mid
			P \in\C\langle X,Y\rangle,
			\ket v\in \ker{\oL}}
		\end{equation}
		is dense in the Hilbert space $\H$.
	\end{hypenum}
	Then, for every $\orho_0 \in \K_d$,
	there exists $\orho_\infty \in \K^1(\H_\oL)\cap \K_d$
	such that $\orho_t \to \orho_\infty$ in trace norm.
\end{theorem}

\subsection{Discussion of the hypotheses}
\label{ssec-hypdisc}
\cref{hyp-wellposed} is an obvious prerequisite to any convergence analysis
and should be checked on its own beforehand. Note that it may implicitly induce additional hypotheses,
depending on the tools used to establish well-posedness -- such as
\cref{lem-existence-semigroup,thm-conservativity-semigroup} from \cref{section-definition-lindblad}
in all examples presented in this article.
\cref{hyp-domain} is a technical assumption,
(which, in particular, implies that the space
$\E_\oL$ in \cref{hyp-density} is well-defined).
Note that $\oG^\dag$ vanishes on $\H_\oL$ so that we always have $\H_\oL \subset D((\oG^\dag)^\infty)$;
moreover $D((\oG^\dag)^\infty)$ is invariant by the semigroup generated by $\oG^\dag$.
The reason for introducing an intermediate subspace $D$ is that, in practice,
we may not want to explicity characterize $D((\oG^\dag)^\infty)$, let alone show that it is stable by $\ob^\dag$:
for instance, in the example developped in \cref{sec-catqubits}, $\oG$ is defined
as the closure of its restriction to a smooth subspace,
and we do not actually need to characterize $D(\oG)$ or $D(\oG^\dag)$.

Intuitively, \Cref{hyp-tightness}
ensures that no mass can be sent to infinity.
In practice, on the examples considered in \cref{sec-catqubits},
this property is established by showing that a suitable observable,
which could be interpreted as an energy operator,
stays bounded along trajectories.

The core of \cref{th-main_convergence} is thus essentially contained
in \cref{hyp-density},
which allows to replace a question of convergence
with an algebraic question of density.
As a result, \cref{th-main_convergence}
changes the nature of the question
rather than simplifying it \emph{per se},
as the density of $\E_\oL$ could
prove no easier to establish on specific examples.
However, the study of multi-photon dissipation processes performed
in \cref{sec-catqubits},
inspired by recent developments in the quantum information literature,
provides an example where establishing \cref{hyp-density} is a viable way to attack the problem
-- and, to the best of our knowledge, the first one for this specific problem.

\subsection{Sketch of the proof and main ideas}
\label{ssec-thm1proofsketch}

The proof of \cref{th-main_convergence} is split into several steps.
In this section,
we break it down into
\cref{lemma-thmweakening}, \cref{thm-main_weak,pt-positivity,pt-final-proof}
to showcase the overall architecture of the proof.
The detailed proof of each step will be provided in
\cref{ssec-proof_mainthm}.
\newline

In a first step, we show that it is enough to obtain a slightly weaker conclusion,
namely that the mass supported in $\H_\oL$ converges to $1$:
\begin{restatable}{lemma}{thmweakening}
	\label{lemma-thmweakening}
	Assume that $\orho_0 \in\K_d$ is such that
	the solution $\orho_t$ of \cref{eq-maineq} starting from $\orho_0$
	satisfies
	$\Tr{\orho_t \projL}\to 1$.
	Then, there exists $\orho_\infty \in \K^1(\H_\oL)\cap \K_d$
	such that $\orho_t \to \orho_\infty$ in trace norm.
\end{restatable}

We then need only prove the following weaker version of \cref{th-main_convergence}:
\begin{theorem}
	\label{thm-main_weak}
    Assume \cref{hyp-wellposed,hyp-domain,hyp-tightness,hyp-density} hold.
    Then, for any $\orho_0\in \K_d$, $\Tr{\orho_t \projL}\to 1$.
\end{theorem}

Note that $\H_\oL$ is absorbing in the following sense:
denoting $(\T_t)_{t\geq 0}$ the
quantum dynamical semigroup on operators
associated to the Lindblad equation \cref{eq-maineq},
and recalling that $\projL$ is the orthogonal projector onto $\H_\oL$,
we have
\(
    \T_t (\projL) \geq \projL
\)
for all $t\geq 0$,
which is straightforwardly deduced from the fact that both $\oL$ and $\ob$ cancel
on $\H_\oL$.
This simple remark suggests drawing inspiration for the proof of \cref{thm-main_weak}
from the classical theory of continuous-time Markov chains with an absorbing state
(however, the actual proofs proposed in the remainder of the paper
will not require any degree of familiarity
with the subject of Markov chains).

Using \cref{hyp-density}, we show that in any finite time,
a strictly positive proportion of the mass
supported on $\H_\oL^\perp$ ends up in $\H_\oL$,
reminiscent of the absorbing state of a Markov chain being reachable
from any other state:
\begin{restatable}{lemma}{lemmapositivity}
	\label{pt-positivity}
	For any $t > 0$,
	$\T_t(\projL)$ has the block-diagonal decomposition
	\begin{equation}
		\T_t(\projL) = \projL + (\Id-\projL) \, \T_t(\projL) \, (\Id-\projL)
	\end{equation}
	with $(\Id-\projL) \, \T_t(\projL) \, (\Id-\projL)$
	definite positive on $\H_\oL^\perp$.
\end{restatable}

We can then combine \cref{pt-positivity} with \cref{hyp-tightness},
which is reminiscent of the condition of a Markov chain not to send mass to infinity,
to prove that when sampling the trajectory at discrete times,
the mass supported on $\H_\oL$ grows at least as fast as a sequence converging arbitrarily close to one:
\begin{restatable}{corollary}{cordiscretetimemass}
	\label{pt-final-proof}
	For every $t_0>0$ and $\epsilon>0$, there exists $\delta \in (0,1)$
	such that for any $n\in\mathbb N$:
	\begin{equation}
		\Tr{\orho_{(n+1) t_0} \projL}
		\geq
		(1-\delta) \Tr{\orho_{n t_0} \projL}
		+ \delta (1-2\epsilon),
	\end{equation}
	hence
	\begin{equation}
		\liminf_{n\rightarrow +\infty} \,
		\Tr{\orho_{n t_0} \projL} \geq 1-2\epsilon.
	\end{equation}
\end{restatable}

Finally, since for every $t\geq 0$ we have $\T_{t}(\projL)\geq \projL$, 
the function $t \mapsto \Tr{\orho_t \projL}$
is increasing,
so that we can extend the conclusion of \cref{pt-final-proof}
to continuous time:
\[
	\forall \epsilon>0, \quad
	\liminf_{t\rightarrow +\infty} \, \Tr{\orho_t \projL}  \geq 1-2\epsilon.
\]
Taking $\epsilon$ arbitrarily close to $0$,
we conclude the proof of \cref{thm-main_weak},
which combined with \cref{lemma-thmweakening} concludes the proof of \cref{th-main_convergence}.

\subsection{Proof of Theorem \ref{th-main_convergence}}
\label{ssec-proof_mainthm}

In \cref{ssec-thm1proofsketch},
the proofs of
\cref{lemma-thmweakening,pt-positivity} as well as \cref{pt-final-proof}
were omitted;
from them, \cref{thm-main_weak} and then \cref{th-main_convergence}
were easily deduced.
In the following subsection, we provide these omitted proofs.

\subsubsection{Proof of Lemma \ref{lemma-thmweakening}}

We recall the statement of \cref{lemma-thmweakening}:

\thmweakening*

\begin{proof}
	For $\orho_0\in\K_d$ such that $\Tr{\orho_t \projL} \rightarrow 1$,
	we
define a time-dependent family of self-adjoint positive operators $\oor_t$
by
	\begin{equation}
		\forall t\geq0, \; \oor_t = \projL \, \orho_t \, \projL.
	\end{equation}
	Note that by construction, $\oor_t\in \K^1(\H_\oL)$ with
	\begin{equation}
		\label{eq-rt_on_hl}
		\Tr{\oor_t} = \Tr{ \projL \, \orho_t \, \projL }
				= \Tr{ \orho_t \, \projL} \xrightarrow[t\rightarrow +\infty]{} 1.
	\end{equation}
	The next step is to prove that these
	operators converge in $\K^1$.
	First, we can show that they are
	non-decreasing in the sense of positive operators:
	\begin{equation}
		\forall t,s \geq 0, \quad \oor_{t+s} \geq \oor_t.
	\end{equation}
	Indeed, let us fix $t,s\geq 0$.
	By definition, both $\oor_{t+s}$ and $\oor_t$ vanish on $\H_\oL^\perp$.
	For $\ket u \in \H_\oL$, we have
	\begin{equation}
		\begin{aligned}
		\bra u \oor_{t+s} \ket u
			&= \bra u \projL \, \orho_{t+s} \, \projL \ket u
		= \bra u \orho_{t+s} \ket u\\
			&= \Tr{\orho_{t+s} \kb{u}}
		= \Tr{\T_{t+s} (\kb u) \rho_0}\\
			&= \Tr{\T_t(\T_s(\kb u))\rho_0}.
		\end{aligned}
	\end{equation}

	Since $\H_\oL \subset \D(\oG^\dag)$ and $\oG^\dag$ cancels on $\H_\oL$,
	we also see that
	\begin{equation}
		\kb u = e^{s\oG^\dag} \, \kb u \, e^{s\oG}.
	\end{equation}
	Moreover, using the integral representation formula given in \cref{eq-lindblad-precise2}
	and the fact that $\T$ is a quantum dynamical semigroup, we obtain
	\begin{align}
    \Id \geq \T_s(\kb u)\geq  e^{s\oG^\dag} \, \kb u \, e^{s\oG}=\kb u
	\end{align}
	and then
	\begin{equation}
		\Tr{\T_t(\T_s(\kb u))\rho_0}
		\geq
		\Tr{\T_t(\kb u)\rho_0}
		=\bra u \oor_{t} \ket u,
	\end{equation}
	which concludes the proof that $\oor_{t+s} \geq \oor_t$.
	Since $\oor_{t+s} - \oor_t$ is a self-adjoint positive operator, we have
	\begin{equation}
		\trnorm{\oor_{t+s}-\oor_t}
		= \Tr{\oor_{t+s} - \oor_t}
		= \Tr{\oor_{t+s}}- \Tr{\oor_t}
		\leq 1- \Tr{\oor_t} \xrightarrow[t\rightarrow+\infty]{}0
	\end{equation}
	since $\left(\Tr{\oor_t}\right)_{t\geq 0}$ is a non-decreasing sequence
	converging to one.
	The family $\left(\oor_t\right)_{t\geq 0}$ is thus a Cauchy sequence in $\K^1$ which
	is complete, this family is thus convergent:
	\begin{equation}
		\oor_{t} \xrightarrow[t\rightarrow +\infty]{} \orho_\infty \in\K^1.
	\end{equation}
	Using \cref{eq-rt_on_hl}, we see that $\Tr{\orho_\infty}=1$ so that
	$\orho_\infty\in\K_d \cap \K^1(\H_\oL)$.
	It remains to show that $\orho_\infty$ is also the limit of $\orho_t$,
	as suggested by our choice of notations;
	note that, since $\orho_\infty$ is supported on $\H_\oL$,
	it is a fixed point of the dynamics.
	Using the trivial decomposition
	\( \Id = \projL + (\Id-\projL) \)
	we can write
	\begin{equation}
	\begin{aligned}
		\orho_{t} - \oor_{t}
		&= \orho_{t} - \projL \, \orho_{t} \, \projL\\
		&= (\Id-\projL) \, \orho_{t} \, (\Id-\projL)
			+ \left( \projL \, \orho_{t} \, (\Id-\projL)
			+ (\Id-\projL) \, \orho_{t} \, \projL \right)
	\end{aligned}
	\end{equation}
	so that
	\begin{equation}
		\label{eq-distrhotrunc}
		\trnorm{\orho_{t} - \oor_{t}}
		\leq
		\trnorm{ (\Id-\projL) \, \orho_{t} \, (\Id-\projL) }
		+ \trnorm{ \projL \, \orho_{t} \, (\Id-\projL) + (\Id-\projL) \, \orho_{t} \, \projL}.
	\end{equation}
	Moreover, for any density operator $\orho\in\K_d$ and any orthogonal projector $\proj$,
	we have that
	$\proj \sqrt{\orho} \in \K^2(\H)$,
	$\sqrt\orho (\Id - \proj) \in \K^2(\H)$
	so that, using the Cauchy-Schwarz inequality:
	\begin{equation}\begin{aligned}
		\trnorm{\proj \orho (\Id-\proj)}
		&\leq \hsnorm{\proj \sqrt{\orho}} \, \hsnorm{\sqrt{\orho} (\Id-\proj)}\\
		&= \trnorm{\proj \orho \proj} \, \trnorm{(\Id-\proj) \orho (\Id-\proj)}\\
		&\leq \trnorm{(\Id-\proj) \orho (\Id-\proj)}.
	\end{aligned}\end{equation}
	Injecting this last inequality into \cref{eq-distrhotrunc}
	we obtain
	\begin{equation}
		\label{eq-orhon__rn}
		\trnorm{\orho_{t} - \oor_{t}} \leq
		3 \, \trnorm{ (\Id-\projL) \, \orho_{t} \, (\Id-\projL) }.
	\end{equation}

	From the initial assumption $\Tr{\orho_t \projL} \rightarrow 1$
	and the fact that $\Tr{\orho_t} = 1$,
	we obtain
	\begin{equation}\begin{aligned}
		\trnorm{ (\Id-\projL) \, \orho_{t} \, (\Id-\projL) }
		&= \Tr{ (\Id-\projL) \, \orho_{t} \, (\Id-\projL) }\\
		&= 1 - \Tr{ \projL \, \orho_t \, \projL}\\
		&= 1 - \Tr{\orho_t \projL} \rightarrow 0
	\end{aligned}\end{equation}
	so that from \cref{eq-orhon__rn} we obtain $\orho_{t} - \oor_{t} \rightarrow 0$
	and can then conclude:
	\begin{equation}
		\trnorm{\orho_t-\orho_\infty}
		\leq \trnorm{\orho_t-\oor_t} + \trnorm{\oor_t-\orho_\infty}
		\xrightarrow[t\rightarrow +\infty]{} 0.
	\end{equation}
\end{proof}

\subsubsection{Proof of Lemma \ref{pt-positivity}}
\label{ssec-thm1proof}
We recall the statement of \cref{pt-positivity}:

\lemmapositivity*

\begin{proof}
Let us consider the operator $\T_t(\projL)$ for some time $t>0$.
Since $\T_t$ is a quantum dynamical semigroup, we get
\begin{align}
    \Id \geq \T_t(\projL) \geq 0.
	\label{eq:ineq_proj_id}
\end{align}
	Since $\H_\oL \subset \D(\oG^\dag)$ and $\oG^\dag$ cancels on $\H_\oL$,
	we also see that
	\begin{equation}
		\projL = e^{t\oG^\dag} \, \projL \, e^{t\oG}.
	\end{equation}
Using the integral representation formula given in \cref{eq-lindblad-precise2},
we obtain
	\begin{align}
    \T_t(\projL)\geq  e^{t\oG^\dag} \, \projL \, e^{t\oG}=\projL.
		\label{eq:evol_projector}
\end{align}
	\cref{eq:ineq_proj_id,eq:evol_projector}
	imply that for any $\ket u\in\H_\oL$, we have $\T_t(\projL) \ket u = \projL\ket u = \ket u$.
	In particular, $\T_t(\projL) - \projL =  (\Id - \projL) \, \T_t \, (\Id - \projL)$
	is a positive self-adjoint operator
	canceling on $\H_\oL$; remains to show that it is positive definite on $\H_\oL^\perp$.

Introducing a Hilbert basis $\left( \ket{v_k}\right)_k$ of $\ker\oL$,
	we see that
	\begin{equation}
		\label{eq-decomp_projl_basisHL}
		\projL = \sum_k \kb{v_k} \otimes \kb0.
	\end{equation}
Moreover, using \cref{hyp-domain},
$\ket{v_k}\otimes \ket0$ belongs to $D\subset \D\left( (\oG^\dag)^\infty\right)$
which is stable under the application of $\ob^\dag$ and the semigroup $(e^{s \oG^\dag})_{s\geq 0}$.
	Therefore, for any $n\in\N$ and $(s_0, \ldots, s_n) \in (\R^+)^{n+1}$, we have
\begin{equation}
		e^{s_n\oG^\dag} \,\ob^\dag  \,
		e^{s_{n-1} \oG^\dag} \,\ob^\dag \ldots
		e^{s_1 \oG^\dag} \,\ob^\dag e^{s_0\oG^\dag} \ket{v_k}\otimes \ket0 \in\H,
\end{equation}
and the above expression is a smooth function of each $s_i$.

This allows us to recursively apply
	the integral representation formula in \cref{eq-lindblad-precise2}:
for any $\ket u\in D(\oG)$ and $n\geq 1$, we find
\begin{align}
    \Bra{u}\T_t(\projL)\Ket{u}
	&= \bra{e^{t\oG} u}  \, \projL \, \ket{e^{t\oG} u}
		+ \kappa \int_0^t \bra{\ob \, e^{(t-s)\oG}u}   \T_s(\projL) \,
						\ket{\ob \, e^{(t-s)\oG} u} ds
		\\
	&\geq
		\kappa \int_0^t \bra{\ob\,e^{(t-s)\oG} u}\T_s(\projL)
		\ket{\ob \, e^{(t-s)\oG}  u }ds
		\\
	&=
		\kappa \int_0^t \bra{u} \, e^{(t-s)\oG^\dag} \, \ob^\dag
		\, \T_s(\projL) \,
		\ob \, e^{(t-s)\oG} \,\ket{ u }ds
		\\
	&\geq
		\kappa^n \int_{0\leq t_0 \leq \ldots \leq t_{n-1} \leq t}
		\bra{ u} \,
		e^{(t-t_{n-1})\oG^\dag} \ob^\dag \, e^{(t_{n-1}-t_{n-2})\oG^\dag}\ldots  \,e^{(t_{1}-t_0)\oG^\dag} \ob^\dag \,e^{t_0\oG^\dag} \notag \\
		& \qquad \qquad \qquad
		 \, \projL \, e^{t_0\oG} \,
		\ob \, e^{(t_{1}-t_{0})\oG} \, \ldots
		\ob \, e^{(t-t_{n-1})\oG}
		\ket{u}
		dt_0\ldots dt_{n-1}\\
	&=
		\kappa^n \; \sum_k \int_{0\leq t_0 \leq \ldots \leq t_{n-1} \leq t}
		dt_0\ldots dt_{n-1} \notag\\
	&\quad\quad\quad
		\left| \bra{ u} \,
		e^{(t-t_{n-1})\oG^\dag} \ob^\dag \, e^{(t_{n-1}-t_{n-2})\oG^\dag}\ldots  \,e^{(t_{1}-t_0)\oG^\dag} \ob^\dag \,e^{t_0\oG^\dag} \ket{v_k} \right|^2.
		\label{eq-ineq_semigroup_rpz}
\end{align}
This last expression depends continuously on $\ket u$
and can thus be extended by density to any $\ket u \in \H$.
In particular, let us consider $\ket u\in \H_\oL^\perp$
and assume that \(\bra u \T_t(\projL)\ket u = 0\);
our goal is to prove that $\ket u = 0$.
Combining \cref{eq-ineq_semigroup_rpz,eq-decomp_projl_basisHL} we obtain
\begin{equation}
	\begin{aligned}
		\forall k, \quad 0 &= \int_{0\leq t_0 \leq \ldots \leq t_{n-1} \leq t}
	dt_0\ldots dt_{n-1}\\
		& \quad \quad
	\left|
		\bra u
		e^{(t-t_{n-1})\oG^\dag} \,\ob^\dag  \,
		e^{(t_{n-1}-t_{n-2})\oG^\dag} \,\ob^\dag \ldots
		e^{(t_{1}-t_{0})\oG^\dag} \,\ob^\dag e^{t_0\oG^\dag} (\ket{v_k}\otimes \ket0)
	\right|^2.
	\end{aligned}
\end{equation}
Since the integrand is a smooth function of $(t_0,\ldots,t_{n-1})$,
we obtain
 \begin{align}
    \Bra{u} e^{(t-t_{n-1}) \oG^\dag} \ob^\dag
	 e^{(t_{n-1}-t_{n-2}) \oG^\dag} \ob^\dag
	 \ldots
	 e^{(t_{1}-t_{0}) \oG^\dag} \ob^\dag \Ket{v_k}\otimes \Ket{0}
	 =0, \quad \forall \,0 \leq t_0 \leq \ldots \leq  t_{n-1}\leq t
 \end{align}
 which is extended to any $\ket v\in\ker\oL$ by linearity.
Applying the change of variables $s_k=t_k-t_{k-1}$ (with $s_n = t-t_{n-1}$ and $s_0=t_0$):
\begin{align}
    \Bra{u} e^{s_n \oG^\dag} \ob^\dag
	e^{s_{n-1} \oG^\dag} \ob^\dag
	\ldots
	e^{s_1 \oG^\dag} \ob^\dag
	e^{s_0 \oG^\dag}
	\Ket{v_i}\otimes \Ket{0} =0, \quad \forall \ket v\in\ker\oL, \forall s_k\geq 0, \,\sum_{k=0}^n s_k = t.
 \end{align}
Taking the partial derivative $\partial_{\alpha}$ with $\alpha=(\alpha_0,\ldots \alpha_n)$ of this function at $(s_k)_{0\leq k\leq n}=(t,0,\ldots, 0)$, we get
\begin{align}
    \label{eq-Gdagbdag}
    \Bra{u}
	(\oG^\dag)^{\alpha_n} \ob^\dag
	\ldots
	(\oG^\dag)^{\alpha_1} \ob^\dag
	(\oG^\dag)^{\alpha_0}
	e^{t\oG^\dag} \Ket{v}\otimes \Ket{0}=0,
	\quad \forall \alpha=(\alpha_0,\ldots \alpha_n) \in\N^n, \forall \ket v\in\ker\oL.
 \end{align}
 Finally, note that for any $\ket v\in\ker\oL$, we have $\oG^\dag \Ket{v}\otimes \Ket{0}=0$ so that
$e^{t\oG^\dag} \Ket{v}\otimes \Ket{0}=\Ket{v}\otimes \Ket{0}$.
By linearity,
\cref{eq-Gdagbdag} can be generalized to any (non-commutative) polynomial:
\begin{align}
    \Bra{u} P(\oG^\dag,\ob^\dag) \Ket{v}\otimes\ket 0=0, \quad \forall \ket v\in \ker\oL,
								\forall P \in \C\langle X,Y\rangle
\end{align}
and thus, applying the density hypothesis \ref{hyp-density}
we conclude that $\ket u=0$.
\end{proof}

\subsubsection{Proof of Corollary \ref{pt-final-proof}}
We recall the statement of \cref{pt-final-proof}:

\cordiscretetimemass*

Let us fix $t_0>0$ and $\epsilon>0$.
For any measurable set $B\subset \R$, we define
$\oP_B := \Id_B(\T_{t_0}(\projL))$
the spectral (or Riesz) projector
of the bounded self-adjoint operator
$\T_{t_0}(\projL)$ on $B$;
in particular, we know
that $\Id \geq \T_{t_0}(\projL) \geq 0$
since $\T_t$ is a quantum dynamical semigroup,
so we will naturally
be interested in the family $(\oP_{[\delta,1]})_{0< \delta \leq 1}$.
We first prove a technical lemma stating that,
up to a small error,
we can use $\oP_{[\delta,1]}$ instead of $\projE$
when we need to apply \cref{hyp-tightness}:
\begin{lemma}
\label{lemma-compat_proj}
    Assume \cref{hyp-tightness} holds.
	Then, for any $\orho_0\in\K_d$ and $\epsilon>0$,
	there exists $\delta>0$ such that
	\[ \forall t>0, \quad \Tr{\rho_t \oP_{[\delta,1]}} \geq 1-2 \epsilon. \]
\end{lemma}

\begin{proof}[Proof of \cref{lemma-compat_proj}]
	Take $\orho_0\in\K_d$, $\epsilon>0$,
	$E$ the subspace defined in \cref{hyp-tightness},
	and denote $\projE$ the orthogonal projector onto $E$.
It is enough to show that, for $\delta>0$ small enough,
	$\oP_{[\delta,1]}\geq \projE-\epsilon\Id$.
	Indeed, we would then have 
	\begin{equation}
		\forall t\geq0, \quad
		\Tr{\orho_{t} \oP_{[\delta,1]}}
			= \Tr{\orho_{t} (\oP_{[\delta,1]}-\projE)}
			+\Tr{\rho_{t} \projE}
			\geq 1-2\epsilon.
	\end{equation}
	$\oP_{[\delta,1]}-\projE$ is a self-adjoint operator which is positive
	on $E^\perp$ (because $\oP_{[\delta,1]}$ is positive);
	we thus need to prove the following:
	\begin{equation}
\label{eq-positivity_proof}
        \exists \delta>0, \quad
		\inf_{\|\ket u\|=1, \ket u \in E}
	    \bra u (\oP_{[\delta,1]}-\projE) \Ket{u}\ \geq  -\epsilon.
	\end{equation}
	For any $\delta>0$ and $\ket u\in E$ with $\norm{\ket u} =1$,
	we have
	$$
	| \bra u (\oP_{[\delta,1]}-\projE) \Ket{u} |
	\leq \norm{(\oP_{[\delta,1]}-\projE) \Ket{u}}
	= \norm{\oP_{[\delta,1]} \ket u  - \ket u}.
	$$
	Moreover, $\T_{t_0}(\projL)$ is self-adjoint
	with $\Id \geq \T_{t_0}(\projL) \geq0$,
	and we showed in
	\cref{pt-positivity}
	that $\T_{t_0}(\projL)>0$ so that $\oP_{\{0\}}=0$ and
	for any $\ket u$ we have
	$\ket u = \oP_{(0,1]} \ket u = \lim_{\delta\rightarrow 0^+} \oP_{[\delta,1]} \ket u$.
	Hence:
	\begin{equation}
		\forall \ket u\in E \textrm{ such that } \| \ket u \| =1, \quad
		\norm{ (\oP_{[\delta,1]}-\projE) \Ket{u} }\xrightarrow[\delta \to 0^+]{} 0.
	\end{equation}
	Since $E$ is finite-dimensional, strong and norm convergences coincide,
	so that the previous property leads to
    \begin{align}
        \label{eq-inequality-proj}
        \sup_{\|\ket u\|=1, \ket u \in E}
	    \|(\oP_{[\delta,1]}-\projE) \Ket{u}\| \xrightarrow[\delta \to 0^+]{} 0
    \end{align}
	from which \cref{eq-positivity_proof} follows.

\end{proof}

We can now conclude the proof of \cref{pt-final-proof}:
\begin{align}
    \Tr{\orho_{(n+1)t_0} \projL}&= \Tr{\orho_{nt_0} \T_{t_0}(\projL)}\\
    &= \Tr{\orho_{nt_0} \projL}
	+  \left( \Tr{\orho_{nt_0} \T_{t_0}(\projL)} -\Tr{\orho_{nt_0} \projL} \right)\\
    &\geq \Tr{\orho_{nt_0} \projL}
	+  \delta \left(\Tr{\orho_{nt_0}\oP_{[\delta,1]}} -\Tr{\orho_{nt_0} \projL} \right)\\
    &\geq \Tr{\orho_{nt_0} \projL}+  \delta \left( 1-2\epsilon -\Tr{\orho_{nt_0} \projL} \right)\\
	&=  (1-\delta) \Tr{\orho_{nt_0} \projL}+  \delta ( 1-2\epsilon )
\end{align}
where we used \cref{lemma-compat_proj}
and the operator inequality
\begin{equation}
	\T_{t_0}(\projL)- \projL \geq  \delta (\oP_{[\delta,1]}-\projL),
\end{equation}
which is a direct consequence of the definition of $\oP_{[\delta,1]}$ as a spectral projector of
$\T_{t_0}(\projL)$ and of the fact that $\oP_{[\delta,1]}- \projL$
vanishes on the support $\H_\oL$ of $\projL$.

\section{Application to engineered multi-photon processes}
\label{sec-catqubits}

We now turn our attention to the application of \cref{th-main_convergence}
to the study of multi-photon dissipation processes
used for the stabilization of cat qubits:
our objective is to show that the theorem applies for the choice of operator
\begin{equation}
	\oL = \oa^k - \alpha^k\Id
\end{equation}
where $\alpha\in \C$, $k\in\N^*$ and $\oa$ is the annihilation operator on $\H_a$.
We are able to prove the following result:
\begin{theorem}
    \label{th-convergence_cat_qubit}
    Let $k\in \N^*$, $\alpha\in\C$ and $\kappa>0$.
	Define $\oL = \oa^k - \alpha^k\Id$,
	$\H_\oL = \ker\oL \otimes \ket0$
	and $\oH = \oL\ob^\dag+\oL^\dag\ob$.
	Then, \cref{eq-maineq} is well-posed and
	for every $\orho_0\in \K_d$,
	denoting $\orho_t$ the solution to \cref{eq-maineq}
	initialized in $\orho_0$:
	\begin{align}
	\recallEq{eq-maineq},
	\end{align}
	there exists $\orho_\infty \in \K_d$ supported on $\H_\oL$
	(\emph{i.e.}, $\orho_\infty \in \K_d \cap \K^1(\H_\oL)$)
	such that
    \begin{align}
        \orho_t \xrightarrow[t\to \infty]{} \orho_\infty.
    \end{align}
\end{theorem}
In order to apply \cref{th-main_convergence},
we have to prove that \cref{hyp-wellposed,hyp-domain,hyp-tightness,hyp-density}
hold.
Without loss of generality%
\footnote{
	For $\alpha\in\C$, defining $\theta_\alpha\in\R$ through
	$\alpha= |\alpha|e^{i\theta_\alpha}$,
	one can replace $\alpha$ by $|\alpha|$ in the Lindblad equation in
\cref{th-convergence_cat_qubit} using the unitary change of frame
$\tilde\orho_t = \oU^\dag \, \orho_t \oU$
with $\oU = e^{i\theta_\alpha ({\oa^\dag \oa} - k\ob^\dag \ob)}$.
}, we will restrict our study to the case $\alpha\in\R$ to alleviate notations.

We first recall a few useful definitions in \cref{sec-cat_functional_setting}.
We then check in \cref{sec-catwellposed}
that \cref{hyp-wellposed} of well-posedness
holds, independently of the values
of $k$ and $\alpha$,
through an application of \cref{lem-existence-semigroup,thm-conservativity-semigroup}.
We give an
explicit characterization of $\ker\oL$ in \cref{sec-charac_kerL},
from which
\cref{hyp-domain}
can be deduced.
The principal difficulties thus lie in checking that
\cref{hyp-tightness,hyp-density} hold.

\cref{hyp-tightness} is established
in \cref{sec-cat_compactness}.
We first formally explain,
in the case $\alpha=0$ for simplicity,
how it can be deduced from the fact that a suitably defined
energy observable remains bounded along trajectories.
We then move on to the general case $\alpha\in\R$
and rigorously justify the previous formal reasoning.

\cref{hyp-density} is established
in \cref{sec-density}.
The case where $\alpha=0$ or $k=1$
is easily treated in \cref{sec-alpha0k1} for
the sake of completeness,
although it is mainly irrelevant for the applications in quantum information
using cat qubits.
The details of its proof are not generalizable to the general case,
but it is already informative to note how, in this simple case,
the density result in $\H = \H_a\otimes\H_b$ required in \cref{hyp-density}
is proved through a simpler density result in $\H_a$ only --
this idea will be the \emph{leitmotiv} of the analysis of the general case.
For $\alpha\in\R$ and $k\geq 2$,
the proof of \cref{hyp-density}
is considerably more involved and requires changing the representation of the Hilbert
space, which is isomorphic to the so-called \emph{Bargmann--Fock} space of holomorphic functions,
in which the density result in \cref{hyp-density} is closely related to a problem of
polynomial approximation known as the \emph{Newman--Shapiro problem}.
We recall the definition of the Bargmann--Fock space and the Newman--Shapiro problem in
\cref{sec-bargmann-fock}.
For pedagogy purposes,
the end of the proof is then split according to the value of $k$,
colloquially referred to as the number of "legs" of the cat qubit in the physics literature.
We first present the case $k=2$ in \cref{sec-density-2}.
The case $k\geq 3$ is similar up to small technical complications, and presented in
\cref{sec-proof-density-gen}.

\subsection{Functional setting}
\label{sec-cat_functional_setting}
The Hilbert space of interest is the two-mode Fock space $\H \equiv \H_a \otimes \H_b$
where $\H_a$ and $\H_b$ are isomorphic to $L^2(\R,\C)$.
We denote by $\oa$ and $\oa^\dag$ (resp. $\ob$ and $\ob^\dag$)
the annihilation and creation operators on
$\H_a$ (resp. $\H_b$);
when working in the full space $\H$,
we will alleviate notations by identifying
$\oa$ with $\oa\otimes\Id_b$
(and similarly $\ob$ with $\Id_a\otimes \ob$).

We now define a few useful subspaces, corresponding to Sobolev
spaces and their equivalents for density operators.

\begin{definition}
Let $p,q\in \R$;
	let $(\ket n)_{n\in\N}$ denote the usual Fock basis of $L^2(\R,\C)$;
	we define the Sobolev spaces
\begin{align*}
    \H_a^{p}&= \left\{
	    \Ket{\psi} = \sum_{n\in \N} \psi_{n} \Ket{n}
	    \quad \big| \quad
	    \sum_{n \in \N} (1+n^{p})|\psi_{n}|^2 < \infty
	    \right\}
			\subset \H_a,\\
    \H^{p,q}&= \left\{
	    \Ket{\psi} = \sum_{n,m\in \N} \psi_{n,m} \Ket{n}\otimes \Ket{m}
	    \quad \big| \quad
	    \sum_{n,m \in \N} (1+n^{p}+ m^{q}) |\psi_{n,m}|^2 < \infty
	    \right\}
			\subset \H
\end{align*}
with their associated inner product
\begin{align*}
    \Braket{u|v}_{\H_a^{p}}&= \sum_{n=0}^\infty (1+n^{p}) \, u_n^* \,v_n, \\
    \Braket{u|v}_{\H^{p,q}}&= \sum_{n,m \geq 0}^\infty (1+n^{p}+ m^{q}) \, {u}_{n,m}^* \, v_{n,m}.
\end{align*}
\end{definition}
With these notations,
we have $\D(\oa) = \D(\oa^\dag) = \H^1_a$
(or $\H^{1,0}$ when considered on the full space $\H$);
similarly, $\D(\ob) = \D(\ob^\dag) = \H^1_b$ (or $\H^{0,1}$).
\newline

We also introduce the linear manifold of states
supported on finitely many Fock states:
\begin{align*}
    \H^f=\left\{ \sum_{0\leq n,m \leq N} u_{n,m} \Ket{n}\otimes \ket{m}
			\; \Big| \;
			N \in \N, \, u_{n,m}\in\C \right\}
\end{align*}
and the corresponding space of trace-class operators supported on $\H^f$:
\begin{align*}
    \K^f=\left\{ \sum_{0\leq n_1,n_2,m_1,m_2\leq N} \rho_{n_1,n_2,m_1,m_2} \ket{n_1}\bra{n_2}\otimes \ket{m_1}\bra{m_2}
			\; \Big| \;
			N\in\N, \, \rho_{n_1,n_2,m_1,m_2} \in \C\right\}.
\end{align*}

For $\alpha \in \R$ and $\kappa >0$, we define the operator
     \begin{equation}
	     \oL=\oa^k-\alpha^k\Id_a
     \end{equation}
     with domain $\D(\oL)=\H_a^k$.
     As usual, we will also use the notation
     $\oL$ as a shorthand for the operator $\oL\otimes\Id_b$ on $\H$,
in which case $\D(\oL) = \H^{k,0}$.
We also recall the notations used in \cref{th-main_convergence}:
\begin{align*}
	\oH=\oL \ob^\dag + \oL^\dag \ob, \qquad \oG=-i\oH - \frac{\kappa}{2} \ob^\dag \ob.
\end{align*}

\subsection{Well-posedness}
\label{sec-catwellposed}
Let us first show that $\oG$, when defined on a suitable domain,
is the generator of a contraction semigroup on $\H$.
Let us initially consider $\oG$ on $\H^{2k,2}$. For every $\ket\varphi,\ket\psi \in \H^{2k,2}$, we have
\begin{equation}
	|\braket{\oG \varphi | \psi}| \leq (\frac{\kappa}{2}+|\alpha|^k +1)\|\ket{\varphi}\|_{\H} \|\ket{\psi}\|_{\H^{2k,2}}
\end{equation}
so that $\H^{2k,2} \subset \D(\oG^\dag)$; as $\H^{2k,2}$ is dense in $\H$,
we deduce that $\oG$ defined on $\H^{2k,2}$ is closable.
We identify $\oG$ with its closure and denote $\D(\oG)$ its domain.
Besides, for every $u\in \H^{2k,2}$
\begin{equation}
	\Re{ \Braket{u|\oG u}}
	= - \tfrac\kappa2 \Braket{\ob u|\ob u}\leq 0
\end{equation}
so that $\oG$ is dissipative and $\D(\oG) \subset \D(\ob)=\H^{0,1}$.
\begin{lemma}
	\label{lemma_maxi_dissi}
	$(\oG,\D(\oG),\H)$ is a maximally dissipative operator.
\end{lemma}
\begin{proof}
	Let us introduce a regularized operator
	$\oG^{\mu}=\oG- \mu \left(({\oa^\dag \oa})^{k}+(\ob^\dag \ob)^{k}\right)$
	with $\mu>0$ a parameter.
	$\oG^{\mu}$
	is a closed operator on the domain $\H^{2k,2k}$
	and the domain of its adjoint is also $\H^{2k,2k}$.
	One can check that $(\oG^{\mu},\H^{2k,2k},\H)$ and its adjoint are dissipative
	and conclude, using Lumer-Philipps theorem \cite{engel-nagel}[Theorem II.3.17], that
	$\oG^\mu$ generates a contraction semigroup on $\H$.
	
	In order to show maximal dissipativity of $\oG$, we will need stronger regularity
	results on $\oG^\mu$.
	To alleviate the computations,
	we first replace the weight $(\Id + (\oa^\dag \oa)^{2k} + (\ob^\dag \ob)^{2k})$
	that defined the scalar product in $\H^{2k,2k}$
	by the equivalent weight $\Id +\left( \frac{\oa^\dag \oa}{k} + \ob^\dag \ob\right)^{2k}$;
	note that this new weight commutes with $\oa^k \ob^\dag + (\oa^\dag)^k \ob$.
	To avoid any confusion, we denote $\tilde \H^{2k,2k}$ this new Hilbert space
	and similarly add a tilde to operators when considered on $\tilde \H^{2k,2k}$.
	
	Let us now show that there exists $\omega>0$,
	independent of $\mu$,
	such that the closure of $(\tilde \oG^{\mu},\H^{4k,4k},\tilde \H^{2k,2k})$
	is the generator of an $\omega$-quasicontraction semigroup.
	Note that for any $\mu >0$,
	$(\tilde \oG,\H^{4k,4k},\tilde \H^{2k,2k})$
	is an infinitesimally bounded perturbation of the operator
	$- \mu \left(({\oa^\dag \oa})^{k}+(\ob^\dag \ob)^{k}\right)$.
	As $\left( - \mu \left(({\oa^\dag \oa})^{k}+(\ob^\dag \ob)^{k}\right),\H^{4k,4k},\tilde \H^{2k,2k} \right)$
	is negative and self-adjoint,
	it is the generator of an analytic semigroup on $\tilde \H^{2k,2k}$;
	thus, $(\oG^\mu,\H^{4k,4k},\tilde \H^{2k,2k})$ is closed
	and the generator of an analytic semigroup
	\cite{engel-nagel}[Theorem III.2.10].

	For $\ket\varphi \in \H^{4k,4k}$, we find
\begin{align*}
	\Re \braket{\tilde \oG^{\mu} \varphi|\varphi}_{\tilde \H^{2k,2k}}&=
	\Re\braket{i {\alpha}^k (\ob+ \ob^\dag) \varphi|\varphi }_{\tilde \H^{2k,2k}}
	- \frac{\kappa}{2} \bra{\varphi} \, \ob^\dag \ob \, \ket{\varphi}_{\tilde \H^{2k,2k}} \\
	&\quad-\mu \bra{\varphi} \, \left( (\oa^\dag \oa)^{k}+(\ob^\dag \ob)^{k} \right) \, \ket{\varphi}_{\tilde \H^{2k,2k}}
\end{align*}
As $\ob(\ob^\dag \ob)^l$ and $\ob^\dag(\ob^\dag \ob)^l$ are infinitesimally bounded with respect to
	$(\ob^\dag \ob)^{l+1}$ for any $l \in \N$,
we can find $\omega>0$ independent of $\mu$ (but depending on $\alpha$)
such that for every $\ket\varphi$ in $\H^{4k,4k}$
\begin{align*}
	\Re \braket{\tilde \oG^{\mu} \varphi| \varphi}_{\tilde \H^{2k,2k}}& \leq \omega \|\varphi\|_{\tilde \H^{2k,2k}}^2.
\end{align*}
Thus, $(\tilde \oG^\mu-\omega, \H^{4k,4k},\tilde \H^{2k,2k})$ is a dissipative operator.
As we already showed that $(\tilde \oG^\mu, \H^{4k,4k},\tilde \H^{2k,2k})$ is the generator of an analytic semigroup, it is the generator of an $\omega$-quasi-contraction semigroup.

Let us now come back to $(\oG,\D(\oG),\H)$ and prove that it is maximally dissipative.
It is enough to show that $(\oG -( \omega+1)\Id ) \H^{4k,4k}$ is dense in $\H$,
	where $\omega$ is the quasi-dissipativity constant previously found which, crucially,
	did not depend on $\mu$.

Let  $\xi \in \tilde \H^{2k,2k}$.
For every $\mu >0$, since $(\tilde \oG^\mu, \H^{4k,4k}, \tilde \H^{2k,2k})$
generates a quasi-contraction semigroup,
there exists $(\ket{\psi_{\mu,n}})_{n\in \N} \in \H^{4k,4k}$
such that
\begin{equation}
(\oG^\mu -( \omega+1)\Id)\ket{\psi_{\mu,n}} \xrightarrow[n \to +\infty]{\tilde \H^{2k,2k}} \ket{\xi}.
\end{equation}
Besides, as $\|(\oG^\mu -( \omega+1)\Id)^{-1}\|_{\tilde \H^{2k,2k}\to \tilde \H^{2k,2k}}\leq 1$,
we can assume $\|\ket{\psi_{\mu,n}}\|_{\tilde \H^{2k,2k}}\leq \|\ket{\xi}\|_{\tilde \H^{2k,2k}}$
without loss of generality.
We get
\begin{equation}
	\|(\oG-( \omega+1)\Id)\ket{\psi_{\mu,n}} - \ket{\xi}\|_{\H}
	\leq \|(\oG-\oG^\mu) \ket{\psi_{\mu,n}}\|_{\H}
	+ \|(\oG^\mu-( \omega+1)\Id)\ket{\psi_{\mu,n}} - \ket\xi\|_{\H}.
	\label{eq:ineg_regul_Gmu}
\end{equation}

We now use that $\frac{1}{\mu}(\oG-\oG^\mu)=\left(({\oa^\dag \oa})^{k}+(\ob^\dag \ob)^{k}\right)$
is a bounded application from $\tilde \H^{2k,2k}$ to $\H$.
\begin{equation}
	\begin{aligned}
	\|(\oG-\oG^\mu) \ket{\psi_{\mu,n}}\|_{\H}
		&\leq \mu \;
		\|\ket{\psi_{\mu,n}}\|_{\tilde \H^{2k,2k}}
		\,
		\|  ({\oa^\dag \oa})^{k}+(\ob^\dag \ob)^{k}  \|_{\tilde \H^{2k,2k} \to \H}
		\\
		&\leq \mu
		\;
		\|\ket\xi\|_{\tilde \H^{2k,2k}}
		\,
		\|  ({\oa^\dag \oa})^{k}+(\ob^\dag \ob)^{k}  \|_{\tilde \H^{2k,2k} \to \H}
	\end{aligned}
\end{equation}
Noting that this last bound no longer depends on $n$,
for any $\epsilon>0$, there exists $\mu>0$ such that for all $n \in \N$,
\begin{align*}
	\|(\oG-\oG^\mu) \ket{\psi_{\mu,n}}\|_{\H}\leq \frac{\epsilon}{2}.
\end{align*}
Then, taking $n$ large enough in
\cref{eq:ineg_regul_Gmu}
leads to $\|(\oG-( \omega+1)\Id)\ket{\psi_{\mu,n}} - \ket{\xi}\|_{\H}\leq \epsilon$.
This shows that  $(\oG-(\omega+1)\Id)\H^{4k,4k}$ is dense (where
$\H^{4k,4k} \subset D(\oG)$); hence, $(\oG,\D(\oG),\H)$ is maximally dissipative.

\end{proof}

Using \cref{lemma_maxi_dissi} and Lumer-Philipps theorem \cite{engel-nagel}[Theorem II.3.15]
we obtain that
$\oG$ is the generator of a semigroup of contraction on $\H$.
Moreover, $\D(\oG)\subset \D(\ob)$ so that \cref{lem-existence-semigroup}
applies: there exists a quantum dynamical semigroup $(\T_t)_{t\geq 0}$ satisfying \cref{eq-Lindblad-precise}.
To establish the conservativity of the minimal semigroup, we take advantage of the following
regularity preservation property of $\oG$:
\begin{lemma}
	\label{lem_stabilityH_2}
	$\H^{2k,2k}$ is $\oG$-admissible, \emph{i.e.}, the restriction of its semigroup to $\H^{2k,2k}$ is a strongly continuous semigroup in $\H^{2k,2k}$ (equipped with its norm).
\end{lemma}
\begin{proof}
	The proof follows closely that of \cref{lemma_maxi_dissi}, we just have to work with more regularity.
	Indeed, we already proved that $(\oG,\H^{4k,4k},\tilde \H^{2k,2k})$ is an $\omega$-quasi-dissipative operator. Thus, it remains to prove that its closure is maximally quasi-dissipative.
	Introducing again a regularized operator
	$\oG^\mu = \oG- \mu \left(({\oa^\dag \oa})^{k}+(\ob^\dag \ob)^{k}\right)$
	but this time on the domain $\H^{6k,6k}$ and with value in $\tilde \H^{4k,4k}$,
	we have a set of closed operators that are maximally $\omega'$-quasi dissipative on
	$\tilde \H^{4k,4k}$ for some constant $\omega'>0$ independent of $\mu$.
	Then, the same argument as in \cref{lemma_maxi_dissi} allows to show that $(\oG -( \omega'+1)\Id ) \H^{6k,6k}$ is dense in $\tilde \H^{2k,2k}$.
	Hence, we proved that the closure of $(\oG,\H^{4k,4k},\tilde \H^{2k,2k})$ is the generator of a quasi-contraction semigroup. By density of $\H^{4k,4k}$ in $\tilde \H^{2k,2}$ for the associated topology, it is a core for $(\oG,\D(\oG),\H)$, hence the restriction of the semigroup generated by $(\oG,\D(\oG),\H)$ to $\H^{2k,2k}$ coincide with the one generated by the closure of $(\oG,\H^{4k,4k},\H^{2k,2k})$.
\end{proof}
The conservativity of the minimal semigroup can now be obtained
by applying \cref{thm-conservativity-semigroup}
with $\oPhi = \oC = \kappa \ob^\dag \ob$
and
$D=\H^{2k,2k}$. Indeed, $\D(\oG)\subset\H^{0,1}=\D(\sqrt{\ob^\dag \ob})$, $\H^{2k,2k}$ is a core for $\sqrt{\ob^\dag \ob}$ and using \cref{lem_stabilityH_2} it is stable by the semigroup generated by $\oG$.
Moreover, for $\lambda$ large enough, the following required inclusions are satisfied
for $k\geq1$:
\begin{align*}
	&\left(\oG -\lambda\Id \right)^{-1}(\H^{2k,2k}) \subset \H^{2k,2k}
	\subset \H^{0,1}=\D(\sqrt{\ob^\dag \ob}),\\
	&\ob \left(\oG -\lambda\Id\right)^{-1}(\H^{2k,2k}) \subset \ob \H^{2k,2k} =\H^{2k,2k-1} \subset  \H^{0,1}=\D(\sqrt{\ob^\dag \ob}).
\end{align*}
\newline

This concludes the proof that \cref{eq-maineq} is well-posed, \emph{i.e.},
\cref{hyp-wellposed} of \cref{th-main_convergence} is satisfied.
We provide a last useful lemma that uses the same tools as above and will be required to establish
\cref{hyp-domain} in the next section:
\begin{lemma}
	\label{lem_stabilityH_l}
	For all $l\in \N$,
	$\H^{2lk,2lk}$ is $\oG^\dag$-admissible.
\end{lemma}
\begin{proof}
	Following the arguments given in the proof of \cref{lem_stabilityH_2}, we get that the closure of $(\oG^\dag,\H^{2(l+1)k,2(l+1)k},\H^{2lk,2lk})$ generates a strongly continuous semigroup. All that remains to show is that it coincides with the restriction to $\H^{2(l+1)k,2(l+1)k}$ of $(e^{t\oG^\dag})_{t\geq 0}$, defined as the adjoint semigroup of $(\oG,\D(\oG),\H)$.
We already know that $\H^{2k,2k} \subset \H^{2k,2} \subset \D(\oG^\dag)$, and it is clear that $\H^{2(l+1)k,2(l+1)k}$ is dense in $\H^{2k,2k}$ for the associated topology. Besides, as the closure of $(\oG^\dag,\H^{2k,2k} ,\H) $ is a maximally dissipative operator it coincides with $(\oG^\dag, \D(\oG^\dag),\H)$, that is $\H^{2k,2k}$ is a core for $(\oG^\dag, \D(\oG^\dag),\H)$, which concludes the proof.
\end{proof}

Let us finally mention that the authors of the recent preprint~\cite{gondolfEnergyPreserving2023}
developed specific tools for the study of the well-posedness
of Lindblad equations where the generator is a polynomial in bosonic annihilation and creation operators,
which may provide another strategy to establish the results of this section.
Since well-posedness could also be easily derived from standard techniques in our case,
we chose to favor a more traditional and self-contained presentation.
We emphasize that the true focus of our study lies in the convergence study,
which is not covered in~\cite{gondolfEnergyPreserving2023},
where in fact only the existence of adherent points is established --
while we are interested in the full characterization of the steady-states and the convergence toward them.

\subsection{Characterization of the kernel of \textbf{\emph{L}}}
\label{sec-charac_kerL}
Recall that, for any%
\footnote{Note that, when $z$ happens to be an integer, the
so-defined coherent state $\ket z$
does \emph{not} coincide with the corresponding Fock state except for $z=0$.
Despite this ambiguity, we keep this notation for sake of consistence with the literature,
and rely on notations to avoid such ambiguities, never using letters such as $k,n,m$
for the complex number denoting a coherent state.}
complex number $z\in\C$, the associated \emph{coherent state}
is defined as
\begin{equation}
	\label{eq-def_coherent_state}
	\ket z = e^{-\frac{|z|^2}2} \sum_{n\in\N} \frac{z^n}{\sqrt{n!}}\ket n
\end{equation}
and is an eigenvector of the annihilation operator $\oa$
associated to the eigenvalue $z$, with
\[ \ker{\oa-z\Id} = \Span{\ket z}.\]

With these definitions, one can easily check that for $\alpha\neq 0$ we have
\begin{align}
	\label{eq-kernel_cat}
	\ker{\oL} = \Span{\ket{\alpha \, e^{\frac{2ir\pi}{k}}}\mid 0\leq r \leq k-1}
\end{align}
whereas for $\alpha=0$ we have
\begin{align}
	\label{eq-kernel_cat_0}
	\ker{\oL} = \Span{\ket0, \ket 1, \ldots \ket{k-1}}.
\end{align}
In both cases, $\dim{\ker{\oL}} = k$.
In particular, for $k=1$, $\ker\oL$ is one-dimensional hence too small
to encode the state of a qubit,
so that we will be mostly interested in the case $k\geq 2$.

It will sometimes be useful to have an orthogonal basis of $\ker\oL$
when $\alpha\neq 0$, since the vectors $\ket{\alpha \, e^{\frac{2ir\pi}{k}}}$
are not orthogonal to each other.
Denoting $\omega = e^{\frac{2i\pi}{k}}$ a $k$-th root of unity,
a straightforward computation shows that the vectors
\begin{equation}
	\ket{\psi_\oL^r} = \sum_{j=0}^{k-1} \omega^{rj}\ket{\alpha \omega^j}, \quad 0\leq r\leq k-1
\end{equation}
are orthogonal to each other:
using \cref{eq-def_coherent_state} to expand $\ket{\psi_\oL^r}$ in the Fock basis for $0\leq r\leq k-1$,
we see that each $\ket{\psi_\oL^r}$ is only supported on the Fock states $\ket n$ satisfying
$n \equiv k-r \mod k$.
We thus have the equivalent expression:
\begin{equation}
	\label{eq-kerL_orthobasis}
	\ker\oL = \Span{ \ket{\psi_\oL^r} \; \mid \; 0\leq r \leq k-1}.
\end{equation}

Next, we define the rotation operator
\begin{equation}
	\oR_\theta = e^{i\theta\oa^\dag\oa}, \quad \theta\in\R
\end{equation}
which acts on coherent states as $\oR_\theta \ket z = \ket{e^{i\theta}z}$.
We see that the previous family satisfies
\begin{equation}
	\label{eq-parity_basis_psir}
	\oR_{\frac{2\pi}k} \ket{\psi_\oL^r} = \omega^{-r} \ket{\psi_\oL^r}.
\end{equation}
For $k=2$, the operator $\oR_{\frac{2\pi}k} = e^{i\pi\oa^\dag\oa} = (-1)^{\oa^\dag\oa}$
is also known as the parity operator.
Using \cref{eq-parity_basis_psir},
we see that \cref{eq-kerL_orthobasis}
provides a decomposition of $\ker\oL$ on the orthogonal eigenspaces of $\oR_{\frac{2\pi}k}$.

Let us now conclude by showing that \cref{hyp-domain} is statisfied. We consider
$$D = \cap_{l\geq 0} \H^{2lk,2lk}.$$
We have that $\ker{\oL}\otimes \ket{0}\subset D \subset \D((\oG^\dag)^\infty)$, $\ob^\dag D \subset D$ and using \cref{lem_stabilityH_l}, $D$ is invariant by the semigroup $(e^{t\oG^\dag})_{t\geq 0}$.

\subsection{Compactness result}
\label{sec-cat_compactness}

The goal of this section is to prove that \cref{hyp-tightness} holds.
The main idea is that we can define a suitable energy operator
$\oV$,
positive self-adjoint with diverging eigenvalues,
so that the quantity $\Tr{\oV \orho_t}$
stays bounded along smooth trajectories $\orho_t$;
choosing the subspace $E$ appearing in
\cref{hyp-tightness} as a sum of eigenspaces of $\oV$
corresponding to its first eigenvalues,
boundedness of $\oV$ along a trajectory implies that the mass supported on
$E$ can be made arbitrarily close to $1$.

For pedagogy purposes, we first sketch how to make this intuition
more precise in the case $\alpha=0$,
before showing how to choose $\oV$ in the general case $\alpha\in\R$
and how to rigorously obtain \cref{hyp-tightness}.

\subsubsection{\texorpdfstring{Main intuition from the case $\alpha=0$}%
{Main intuition from the case \unichar{"1D736}=0}}
\label{sec-compact_intuition_alpha0}

Let us consider $\alpha=0$ hence $\oL = \oa^k$.
Considering that the Hamiltonian
\[
	\oH = \oL \ob^\dag + \oL^\dag \ob
	= \oa^k \ob^\dag + \oa^{\dag k} \ob
\]
appearing in
\cref{eq-maineq}
describes an interaction process where $k$ photons in mode $a$
can be transformed into one photon in mode $b$ and vice versa,
it is natural to define a rescaled full energy of the system
as
\begin{equation}
	\label{eq-energy_rescaled_nopower}
	\oV = \frac{\oa^\dag \oa}k + \ob^\dag\ob.
\end{equation}
Formally, one has
$[\oH, \oV] = 0$
hence, in the Heisenberg picture (see \cref{eq-formal-lindblad_adjoint}),
we have
\begin{equation}
	\label{eq-dotV_nopower}
	\begin{aligned}
		\L^*(\oV) &= i [ \oH, \oV] + \kappa \, D^*[\ob](\oV) \\
			&= \kappa \, D^*[\ob](\oV) \\
			&= \kappa \, D^*[\ob](\ob^\dag \ob) \\
			&= - \kappa \, \ob^\dag \ob.
	\end{aligned}
\end{equation}

In particular, temporarily dismissing any regularity consideration for the sake of intuition,
we have
\begin{equation}
	\label{eq-alpha0_Vdot}
	\frac d{dt} \Tr{\oV\orho_t}
	= \Tr{\L^*(\oV)\orho_t}
	= -k \Tr{\ob^\dag\ob \orho_t} \leq 0
\end{equation}
hence, assuming $\Tr{\oV\orho_0} < \infty$,
\begin{equation}
	\Tr{\oV\orho_t} \leq \Tr{\oV\orho_0}
\end{equation}
so that $\oV$ stays bounded along trajectories $\orho_t$.
Moreover, $\oV$ is diagonal in the Fock basis;
its eigenvalues are all the combination
\begin{equation}
	d = \frac nk + m, \quad n,m\in\N
\end{equation}
and the corresponding eigenspaces are finite-dimensional, with
\begin{equation}
	\forall n,m\in \N, \quad
		\ker{\oV - ( \tfrac nk + m ) \Id}
			= \Span{\ket{n'} \otimes \ket{m'}
				\; \mid \;
				\tfrac{n'}k + m' = \tfrac nk + m}.
\end{equation}
For any $\orho_0$  satisfying $\Tr{\oV\orho_0}<\infty$
and any $\epsilon >0$,
defining a finite-dimensional subspace $E$ as
\begin{equation}
	\begin{aligned}
		E &=
			\oplus_{\tfrac nk + m \leq \frac 1\epsilon}
			\ker{\oV - ( \tfrac nk + m ) \Id}\\
		&= \Span{ \ket{n} \otimes \ket{m}
			\; \mid \;
			\tfrac{n}k + m \leq \tfrac1\epsilon},
	\end{aligned}
\end{equation}
denoting $\projE$ the orthogonal projector onto $E$
and exploiting the fact that any density operator $\orho$
can be decomposed in the Fock basis as
\begin{equation}
	\orho =
		\sum_{n_1,n_2,m_1,m_2\in\N}
			\orho_{n_1,n_2,m_1,m_2}
			\, \left( \ket{n_1} \otimes \ket{m_1} \right)
			\, \left( \bra{n_2} \otimes \bra{m_2} \right),
\end{equation}
we obtain
\begin{equation}
	\begin{aligned}
	\forall t\geq 0, \quad \Tr{\projE \orho_t}
		&= \sum_{\tfrac nk+m \leq \tfrac 1\epsilon}
			\orho_{n,n,m,m}(t)\\
		&= \Tr{\orho_t} - \sum_{\tfrac nk+m > \tfrac 1\epsilon}
			\orho_{n,n,m,m}(t)\\
		&= 1 - \sum_{\tfrac nk+m > \tfrac 1\epsilon}
			\frac{ \left( \tfrac nk + m \right) \orho_{n,n,m,m}(t)}
				{\tfrac nk + m}\\
		&\geq 1 - \epsilon \left( \sum_{\tfrac nk+m > \tfrac 1\epsilon}
			\left( \tfrac nk + m \right) \orho_{n,n,m,m}(t) \right) \\
		&\geq 1 - \epsilon \left( \sum_{n,m\in \N}
			\left( \tfrac nk + m \right) \orho_{n,n,m,m}(t) \right)\\
		&= 1 - \epsilon \Tr{\oV \orho_t}\\
		&\geq 1 - \epsilon \Tr{\oV \orho_0}
	\end{aligned}
\end{equation}
which would formally conclude the proof of \cref{hyp-tightness}.
The remainder of this section will be consecrated to
both extending this proof to the general case $\alpha\in\R$ and
rigorously formalizing it.
In particular,
we mention that
the reader familiar with Lyapunov theory could object at this point
that, from the positivity of $\oV$ and \cref{eq-alpha0_Vdot},
the function $t\mapsto \Tr{\oV \orho_t}$ is a Lyapunov function of the system
and one could try to rigorously formalize a convergence proof based on a
direct application of
LaSalle's invariance principle, instead of resorting to \cref{th-main_convergence};
and indeed,
if one could find a similar
Lyapunov function in the case $\alpha\neq 0$,
it could possibly be a drastically simpler strategy.
However, we failed to find such a Lyapunov function
generalizing
\cref{eq-alpha0_Vdot}
when $\alpha\neq 0$.

\subsubsection{A formal \textit{a priori} estimate}
Let us now consider arbitrary $k\in\N$ and $\alpha\in\R$.
We momentarily keep working with formal manipulations
to find an adapted energy operator $\oV$ bounded along trajectories,
in the sense that all operator equalities below should be understood as
applied on a subspace of sufficiently smooth vectors, for instance $\H^f$.
A rigorous justification of the so-obtained estimate will later be provided in
\cref{sec-Justification-of-apriori}.

Using again
\(
\left[ \oa^k \ob^\dag + \oa^{\dag k} \ob, \,
	\frac{\oa^\dag \oa}k + \ob^\dag \ob \right] = 0
\)
and trying to extend \cref{eq-energy_rescaled_nopower,eq-dotV_nopower},
we find
\begin{equation}
	\label{eq-fail_Vdot}
	\begin{aligned}
	\L^*\left( \frac{\oa^\dag \oa}k + \ob^\dag \ob \right)
		&= - \kappa \, \ob^\dag \ob
			-i\alpha^k \,
				\left[ \ob + \ob^\dag, \frac{\oa^\dag \oa}k + \ob^\dag \ob \right]\\
		&= - \kappa \, \ob^\dag \ob
			-i \alpha^k (\ob - \ob^\dag).
	\end{aligned}
\end{equation}
The additional second term proportional to $\ob - \ob^\dag$
is neither negative
nor controllable by an operator of the form
\(
- \mu \left( \frac{\oa^\dag \oa}k + \ob^\dag \ob\right) + \gamma \Id
\)
for some constants $\mu,\gamma >0$, so that overall
\cref{eq-fail_Vdot}
is not immediately usable to bound the evolution of
$ \Tr{ \left(\frac{\oa^\dag \oa}k + \ob^\dag \ob\right) \orho_t}$.
On the other hand, $\ob - \ob^\dag$ is infinitesimally bounded by $\ob^\dag \ob$;
if the right hand side of \cref{eq-fail_Vdot}
contained a term proportionnal to $\oa^\dag \oa$ with a negative coefficient,
we could thus use a Gronwäll argument to
conclude that $\frac{\oa^\dag \oa}k + \ob^\dag \ob$ stays bounded along trajectories.
We can thus try to modify the operator $\frac{\oa^\dag \oa}k + \ob^\dag \ob$
appearing on the left-hand side to make such a term proportionnal to $\oa^\dag \oa$ appear
on the right-hand side.

Inspired by classical hypocoercivity ideas,
we introduce the operator
\begin{align}
	\oW&=  \mathcal{L}^*\left(\frac{\oa^\dag \oa}k \right)
	= \frac ik \left[ \oa^k \ob^\dag + \oa^{\dag k} \ob, \, \oa^\dag \oa \right]
	= i  \left( \oa^k \ob^\dag - \oa^{\dag k} \ob \right)
\end{align}
and choose the $k$-th power of the previous rescaled energy as our candidate operator
$\oV$:
\begin{equation}
	\oV = \left( \frac{\oa^\dag \oa}k + \ob^\dag \ob \right)^k.
\end{equation}
The intuition behind the introduction of the $k$-th power can be understood
as to make $\oW$ infinitesimally bounded by $\oV$.
Hence, it is equivalent to obtain that $\Tr{\oV \orho_t}$
is bounded along trajectories or that $\Tr{(\oV+\mu \oW) \orho_t}$
is for some small parameter $\mu>0$.

Our main objective is thus to obtain (formally, with a rigorous justification in the next section)
the following result:
\begin{proposition}
	\label{prop-Vbounded}
	Assume $\orho_0 \in \K^f \cap \K_d$.
	Then, there exists a constant $C>0$ depending on $\orho_0$
	such that:
	\begin{equation}
		\forall t\geq 0, \quad \Tr{\oV \orho_t} \leq C
	\end{equation}
	with $\oV = \left( \frac{\oa^\dag \oa}k + \ob^\dag \ob \right)^k$.
\end{proposition}
The strategy to establish \cref{prop-Vbounded}
consists in proving \cref{lemma-hypcoercive} below, which is equivalent since $\oW$ is
infinitesimally bounded by $\oV$:
\begin{lemma}
	\label{lemma-hypcoercive}
	Assume $\orho_0 \in \K^f \cap \K_d$.
	Then, there exists a constant $C>0$ depending on $\orho_0$
	and $\mu>0$
	such that:
	\begin{equation}
		\forall t\geq 0, \quad \Tr{\left( \oV + \mu\oW \right) \orho_t} \leq C
	\end{equation}
	with $\oV = \left( \frac{\oa^\dag \oa}k + \ob^\dag \ob \right)^k$
	and $\oW = i \left( \oa^k \ob^\dag - \oa^{\dag k} \ob \right)$.
\end{lemma}

Let us now estimate the evolution of $\oV$ and $\oW$ along trajectories.
Using that $\oV$ commutes with $\oa^k\ob^\dag + \oa^{\dag k} \ob$,
we have
    \begin{align}
        \L^*(\oV)=
        &i[\oH,\oV]+ \kappa D^*[\ob](\oV)\\
        &=-i \alpha^k [\ob+\ob^\dag, \oV] + \kappa D^*[\ob](\oV).
	    \label{eq-Vdot_formal}
    \end{align}
Since $\oa^\dag \oa$ and $\ob^\dag \ob$ commute,
we have
\begin{equation}
	\label{eq-binom_V}
	\oV = \sum_{j=0}^k \binom kj \left( \frac{\oa^\dag\oa}k \right)^{k-j}
				\left( \ob^\dag \ob \right)^j,
\end{equation}
and we can explicitly compute each term in \cref{eq-Vdot_formal}:

    \begin{align}
        [\ob+\ob^\dag, \oV]
	    &= \sum_{j=0}^k  \binom{k}{j}
		\left[\ob+\ob^\dag, \left(\ob^\dag \ob \right) ^j \right]
			\left( \frac{\oa^\dag \oa}{k}\right)^{k-j}\\
	    &= \sum_{j=1}^k  \binom{k}{j}
		\left[\ob+\ob^\dag, \left(\ob^\dag \ob \right) ^j \right]
			\left( \frac{\oa^\dag \oa}{k}\right)^{k-j}.
			\label{eq-commut_qb_V}
    \end{align}
    One can easily check that in $\H^f$, we have
    \begin{equation}
	    \label{eq-commute_power_nb}
    \begin{aligned}
	    \ob \, (\ob^\dag \ob)^j &= (\ob^\dag \ob+\Id)^j \, \ob,\\
	\ob^\dag \, (\ob^\dag \ob)^j &= (\ob^\dag \ob-\Id)^j \, \ob^\dag
    \end{aligned}
    \end{equation}
    so that
    \begin{equation}
	    \left[ \ob + \ob^\dag, (\ob^\dag\ob)^j) \right]
		= \left( (\ob^\dag \ob + \Id)^j - (\ob^\dag\ob)^j\right)\ob
			+ \left( (\ob^\dag \ob - \Id)^j - (\ob^\dag\ob)^j\right)\ob^\dag,
    \end{equation}
	hence all terms appearing in the sum \cref{eq-commut_qb_V}
	can be expanded as linear combinations of operators of the form:
    \begin{align}
        \label{eq-monomial_V}
        c_{j_1,j_2}(\oa^\dag \oa)^{j_1} (\ob^\dag \ob)^{j_2} \ob,
	    \quad c'_{j_1,j_2}(\oa^\dag \oa)^{j_1} (\ob^\dag \ob)^{j_2} \ob^\dag ,
	    \qquad \text{with } j_1+j_2 \leq k-1.
    \end{align}
    Recall that $\ob$ and $\ob^\dag$ are easily shown to be infinitesimally bounded
    with respect to $\ob^\dag \ob$,
    using the Cauchy-Schwarz identity and the fact that $\D(\ob^\dag \ob) \subset \D(\ob)$:
    \begin{equation}
	    \begin{aligned}
	    \forall \epsilon>0,
	    \forall \ket u \in \D(\ob^\dag\ob) = \H^{0,2},
	    \quad
		\norm{\ob \ket u}^2
		    &= \bra u \ob^\dag \ob \ket u
		\leq \norm{u} \norm{\ob^\dag \ob \ket u}\\
		    &\leq \frac 1{4\epsilon} \norm{\ket u}^2 + \epsilon \,\norm{\ob^\dag \ob\ket u}^2\\
		\norm{\ob^\dag \ket u}^2
		    &= \bra u \ob \ob^\dag \ket u
		    = \bra u (\ob^\dag \ob+\Id) \ket u\\
		    &= \norm{u}^2 + \norm{ \ob \ket u}^2.
	    \end{aligned}
    \end{equation}
	    Hence, all terms of the form given in
	    \cref{eq-monomial_V}
	    are infinitesimally bounded with respect to
	    $(\ob^\dag \ob)^k + (\oa^\dag \oa)^k$,
	    and in turn the first term
	    $-i \alpha^k [\ob+\ob^\dag, \oV]$
	    appearing in \cref{eq-Vdot_formal}
	    is infinitesimally bounded with respect to $\oV$.

    Let us now compute the second term in \cref{eq-Vdot_formal}.
Using again \cref{eq-binom_V,eq-commute_power_nb}, we find:
    \begin{align}
        D^*[\ob](\oV)
	    &=\sum_{j=0}^k  \binom{k}{j}
		\left( \frac{\oa^\dag \oa}k\right)^{k-j}
		D^*[\ob]\left((\ob^\dag \ob)^j\right)\\
	    &=\sum_{j=1}^k  \binom{k}{j}
		\left( \frac{\oa^\dag \oa}k\right)^{k-j}
		D^*[\ob]\left((\ob^\dag \ob)^j\right)\\
	    &=\frac12 \, \sum_{j=1}^k  \binom{k}{j}
		\left( \frac{\oa^\dag \oa}k\right)^{k-j}
		\left( \ob^\dag \left[(\ob^\dag \ob)^j,\ob\right]
			+ \left[ \ob^\dag, (\ob^\dag \ob)^j\right] \ob \right)\\
        &=-\sum_{j=1}^k  \binom{k}{j}
		\left( \frac{\oa^\dag \oa}k\right)^{k-j}
		\, \ob^\dag \ob \,
	    \left( (\ob^\dag \ob)^{j}- (\ob^\dag \ob-\Id)^j \right)\\
	    &=- \ob^\dag \ob \, \left( (\ob^\dag \ob)^{k}- (\ob^\dag \ob-\Id)^k \right)
	    -\sum_{j=1}^{k-1}  \binom{k}{j}
		\left( \frac{\oa^\dag \oa}k\right)^{k-j}
		\, \ob^\dag \ob \,
	    \left( (\ob^\dag \ob)^{j}- (\ob^\dag \ob-\Id)^j \right)\\
	    &=	- k \, (\ob^\dag \ob)^k
	    + \sum_{\substack{1\leq j_1\leq k-1\\ 0\leq j_2 \leq k-1}}  d_{j_1,j_2} (\oa^\dag \oa)^{j_1}(\ob^\dag \ob)^{j_2}
	    \label{eq-fin_computation_DbV}
    \end{align}
    for some coefficients $d_{j_1,j_2}$;
    note that all terms in the sum to the right of \cref{eq-fin_computation_DbV}
    are infinitesimally bounded with respect to $\oV$.

    All in all, we can thus write
    \begin{equation}
	    \label{eq-Vdot_leading_term}
	    \L^*(\oV) = -k \, (\ob^\dag \ob)^k + \oR
    \end{equation}
    for some (self-adjoint) operator $\oR$ infinitesimally bounded with respect to $\oV$.

    Now, performing the same computations for $\oW$ we find:
    \begin{align}
	    \L^*(\oW)
	    &= i [\oH, \oW] + \kappa D^*[\ob](\oW)\\
	    &= - \left[ \oa^k \ob^\dag + \oa^{\dag k} \ob, \,
			\oa^k \ob^\dag - \oa^{\dag k} \ob \right]
		+ \alpha^k\, \left[ \ob + \ob^\dag, \, \oa^k \ob^\dag - \oa^{\dag k} \ob \right]
		-\frac\kappa2 \oW\\
	    &= -2 \left[ \oa^{\dag k}\ob, \, \oa^k \ob^\dag\right]
		+ \alpha^k (\oa^k + \oa^{\dag k}) - \frac\kappa2 \oW\\
	    &= -2 \, \oa^{\dag k} \oa^k
		+ 2 \left[ \oa^{\dag k}, \oa^k\right] \, \ob \ob^\dag
		+ \alpha^k (\oa^k + \oa^{\dag k}) - \frac\kappa2 \oW.
    \end{align}
    Iterating over the canonical relationship $[\oa,\oa^\dag]=\Id$
    we find that
    $(\oa^\dag \oa)^k - \oa^{\dag k} \oa^k$
    can be written as a linear combination of the operators $\left((\oa^\dag \oa)^j\right)_{0\leq j\leq k-1}$
    and is thus infinitesimally bounded with respect to $(\oa^\dag \oa)^k$.
    Moreover, a direct computation yields
    \begin{equation}
	    [\oa^{\dag k}, \oa^k]
	    =  (\oa^\dag\oa) \, (\oa^\dag\oa - \Id)^+ \, \ldots (\oa^\dag\oa - (k-1)\Id)^+
		\; - \;
	    (\oa^\dag\oa+\Id) \, \ldots (\oa^\dag\oa + k\Id)
    \end{equation}
    with $x\mapsto x^+$ defined as $x^+=x\,\Id_{[0,+\infty[}(x)$.
    Using this formula and the positivity of $\ob\ob^\dag$,
    we see that $[\oa^{\dag k}, \oa^k] \, \ob\ob^\dag$ is a self-adjoint negative operator.
From the above and similarly to \cref{eq-Vdot_leading_term},
    we can thus write
    \begin{equation}
	    \label{eq-Wdot_leading_term}
	    \L^*(\oW) \leq -2 (\oa^\dag \oa)^k + \oR
    \end{equation}
    with $\oR$ a (self-adjoint) operator infinitesimally bounded with respect to $\oV$.
    Finally, for any $\mu>0$, we deduce from
    \cref{eq-Vdot_leading_term,eq-Wdot_leading_term}:
    \begin{align}
	    \label{eq-Vhypo_dot_1}
	    \L^*(\oV + \mu \oW) &=
		-2 \mu (\oa^\dag \oa)^k
		- k (\ob^\dag \ob)^k + \oR
    \end{align}
    with $\oR$ infinitesimally bounded with respect to $\oV$.
    We emphasize that the operators denoted $\oR$ in
    \cref{eq-Vdot_leading_term,eq-Wdot_leading_term,eq-Vhypo_dot_1}
    do not coincide, the notation $\oR$ being used as a mere shorthand for any operator
     infinitesimally bounded with respect to $\oV$.
     Finally, since $\oV - \left( \left( \frac{\oa^\dag \oa}k \right)^k + (\ob^\dag \ob)^k\right)$
     is infinitesimally bounded with respect to $\oV$,
     we deduce from \cref{eq-Vhypo_dot_1} that,
     for $\mu>0$ small enough,
     there exists two constants $C_1, C_2>0$
     such that the following operator inequality holds:
     \begin{equation}
		\label{eq-inequality_bounded_pot_informal}
	     \L^*(\oV + \mu \oW) \leq C_1 \, \Id - C_2 ( \oV + \mu \oW ).
     \end{equation}

     When applying the previous inequality along a given trajectory,
     formally assuming that $\orho_t$ is smooth enough at all times,
     we find
     \begin{equation}
	     \frac d{dt} \Tr{(\oV + \mu \oW)\orho_t}
		= \Tr{\L^*(\oV + \mu \oW) \orho_t}
		\leq C_1 - C_2 \Tr{ (\oV + \mu \oW)\orho_t}
     \end{equation}
     from which we can deduce
     \begin{equation}
	     \forall t\geq 0, \quad
	     \Tr{(\oV + \mu \oW)\orho_t}
	     \leq
	     \tfrac{C_1}{C_2} + \left( \Tr{(\oV + \mu \oW)\orho_0} - \tfrac{C_1}{C_2}\right) e^{-C_2 t}.
     \end{equation}

     This formally concludes the proof of the inequality in \cref{lemma-hypcoercive}
     with
     \begin{equation}
	     C = \max\left(\tfrac{C_1}{C_2}, \Tr{(\oV + \mu \oW)\orho_0}\right).
     \end{equation}
     To finish the proof of \cref{lemma-hypcoercive},
     it remains to show that the above formal manipulations
     can be made rigorous when $\orho_0 \in \H^f\cap \H_d$,
     which is the object of \cref{sec-Justification-of-apriori}.


\subsubsection{Justification of the estimate}
\label{sec-Justification-of-apriori}

In this section, we assume $\Tr{\orho_0\oV} < \infty$ (which covers in particular
the case $\orho_0 \in \K^f$) and show that
for 
$t \geq 0$,
$\Tr{\S_t(\orho_0) \oV}$ is well-defined and satisfies
\begin{equation}
\sup_{t\geq 0} \Tr{\S_t(\orho_0) \oV}<\infty.
\end{equation}

First note that $\oV+\mu \oW$ is a self-adjoint operator on $\D(\oV)= \H^{2k,2k}$. As $\oV$ is a positive operator and $\oW$ is infinitesimally bounded with respect to $\oV$, the operator $\oV+\mu \oW$ is bounded from below.
Thus, we can choose $\lambda>0$ big enough
so that $\oX= \oV+\mu \oW + \lambda \Id$ is a positive operator. Using $\mathcal{L^*}(\Id)=0$,
where the equality is understood as an equality of quadratic forms on $\H^f$,
we obtain from \cref{eq-inequality_bounded_pot_informal} (with the slight modification $C_1 \to C_1+\lambda$)
    \begin{align}
        \label{eq-inequality_bounded_pot}
        \braket{\oG u|\oX u}+\braket{\oX u|\oG u} +\braket{\oX ^{1/2}\ob u|\oX ^{1/2}\ob u}
        &\leq \bra{u}C_1- C_2 \oX \ket{u},\qquad \ket{u} \in \H^f.
    \end{align}
Let us remark that $\oG $ is relatively bounded with respect to $\oV$ thus also with respect to $\oX $.
    Following \cite{fagnolaExistenceStationaryStates2001}[Section IV], we introduce the Yosida approximations
    \begin{align}
        \oG ^{(n)}=n\oG (n+\oX )^{-1}, \qquad  \ob ^{(n)}=n\ob (n+\oX )^{-1},
    \end{align}
    as well as the bounded linear operator $B(\H) \to B(\H)$
    \begin{equation}
	    \mathcal{L}^{(n)*}(\ox)=\oG ^{(n)\dag}\ox+ \ob^{(n)\dag}\ox\ob ^{(n)}+\ox\oG ^{(n)},
	    				\quad \ox\in B(\H).
    \end{equation}
    As the generator $\mathcal{L}^{(n)*}$ is bounded, we can use Lindblad celebrated result \cite{lindbladGeneratorsQuantumDynamical1976} and define $(\T^{(n)}_t)_{t\geq 0}$ the associated uniformly continuous semigroup on $B(\H)$ and $(\S_t^{(n)})_{t\geq 0}$ the corresponding predual semigroup on $\K^1(\H)$.
    \begin{lemma}[\cite{fagnolaExistenceStationaryStates2001} Lemma IV.3]
        For any $\ket{u}\in \D(\oX )$, we have
        \begin{align}
            \oG ^{(n)}\Ket{u} \xrightarrow[n\to \infty]{}\oG  \Ket{u},\qquad \ob ^{(n)}\Ket{u} \xrightarrow[n\to \infty]{}\ob  \Ket{u}
        \end{align}
        Besides, $\S_t^{(n)}$ converges strongly to $\S_t$ uniformly for $t$ in a bounded interval.
    \end{lemma}
Let us now introduce
    \begin{align*}
        \oX ^{(n)}=n\oX (n+\oX )^{-1} \geq   \tilde \oX ^{(n)}=n^2\oX (n+\oX )^{-2}.
    \end{align*}
    Using that $\H^f$ is a core for $\oX$, we can extend \cref{eq-inequality_bounded_pot} to $\D(\oX )$:
    \begin{align}
	    \Braket{\oG u|\oX u}+ \braket{\oX ^{1/2}\ob  u|\oX ^{1/2}\ob  u}+\Braket{\oX u|\oG u}\leq \Bra{u} C_1-C_2 \oX \Ket{u},\qquad \ket u \in \D(\oX).
    \end{align}
    Applying the previous inequality to $n(n+\oX )^{-1}\Ket{u}$ for any $\ket{u}\in \H$ gives
    \begin{align}
        \mathcal{L}^{(n)*}(\oX ^{(n)})\leq C_1-C_2 \tilde \oX ^{(n)}.
    \end{align}
Thus, one has
    \begin{align*}
        \frac{d}{dt}\left( \T_t^{(n)}(\oX ^{(n)})-C_1 t +C_2 \int_0^t \T_s^{(n)}(\tilde \oX ^{(n)})ds \right)
        &= \mathcal{L}^{(n)*}(\T_t^{(n)}(\oX ^{(n)}))- C_1 \Id +C_2 \T_t^{(n)}(\tilde \oX ^{(n)})\\
        &= \T_t^{(n)}\left(\mathcal{L}^{(n)*}(\oX ^{(n)})- C_1\Id +C_2 \tilde \oX ^{(n)}\right)\leq 0.
    \end{align*}
    As a consequence
    \begin{align}
        \label{eq-integral_inequality_Xn}
        \T_t^{(n)}(\oX ^{(n)})+ C_2 \int_0^t \T_s^{(n)}(\tilde \oX ^{(n)})ds \leq \oX ^{(n)}+ C_1 t.
    \end{align}

    Let us now regularize once more the operator $\oX$ by spectral truncation. For a given Borel set $B$ and a self-adjoint operator $\oS$, we denote $\Id_{B}(\oS)$ the associated spectral (or Riesz) projector.
	For a given $r>0$, we introduce the bounded operator sequences
    \begin{align*}
        \oX ^{(n)}\wedge r = \oX ^{(n)}\Id_{[0,r]}(\oX ^{(n)})+ r\Id_{(r,+\infty)}(\oX ^{(n)})  \leq \oX ^{(n)},\\
		\tilde \oX ^{(n)}\wedge r = \tilde \oX ^{(n)}\Id_{[0,r]}(\tilde \oX ^{(n)})+ r\Id_{(r,+\infty)}(\tilde \oX ^{(n)})  \leq \tilde \oX ^{(n)}.
    \end{align*}
    From \cref{eq-integral_inequality_Xn}, we get for every $\ket{u} \in \D(\oX )$
    \begin{align*}
        \Tr{\S_t^{(n)}(\Ket{u}\Bra{u})(\oX ^{(n)}\wedge r)+ C_2 \int_0^t \S_{s}^{(n)}(\Ket{u}\Bra{u})(\tilde \oX ^{(n)}\wedge r)ds}\leq \bra{u}\oX +C_1 t\ \ket{u}.
    \end{align*}
    Using the uniform in time strong convergence of $(\S_{s}^{(n)})_{0\leq s \leq t}$, we get
    \begin{align*}
        \Tr{\left(\S_t^{(n)}(\Ket{u}\Bra{u})- \S_t(\Ket{u}\Bra{u})\right)(\oX ^{(n)}\wedge r)}\xrightarrow[n\to \infty]{} 0,\\
        \Tr{\int_0^t \left(\S_{s}^{(n)}(\Ket{u}\Bra{u})- \S_{s}(\Ket{u}\Bra{u})\right)(\tilde \oX ^{(n)}\wedge r)ds}\xrightarrow[n\to \infty]{} 0.
    \end{align*}
    Then, for any $\ket{v}\in \H$, we decompose $\ket{v}$ according to the spectral density $d\sigma_X$ of $\oX $:
    \begin{align*}
        \Ket{v}= \int_0^\infty \Ket{\hat{v}(\lambda)} d\sigma_X(\lambda),
    \end{align*}
    with $\hat{v}\in L^2([0,\infty],\H, d\sigma_X)$.
    By Lebesgue dominated convergence, we get
    \begin{equation*}
        (\oX ^{(n)}\wedge r) \ket{v}
	    = \int_0^\infty \min \{\frac{n\lambda}{n+\lambda},r\}\Ket{\hat{v}(\lambda)} d\sigma_X(\lambda) \xrightarrow[n\to \infty]{} \int_0^\infty \min\{ \lambda,r\}\Ket{\hat{v}(\lambda)} d\sigma_X(\lambda)=(\oX \wedge r)\ket{v}.
    \end{equation*}
    Similarly, $\tilde \oX ^{(n)}\wedge r$ converges strongly toward $\oX \wedge r$. Thus, for every $\ket{u} \in \D(\oX)$,
    \begin{equation}
        \Tr{\S_t(\Ket{u}\Bra{u})(\oX \wedge r)
        + C_2 \int_0^t \S_{s}(\Ket{u}\Bra{u})( \oX \wedge r)ds}
        \leq \bra{u}\oX +C_1 t\ \ket{u}.
    \end{equation}
We can extend this inequality to $\ket u \in \D(\oX^{1/2})$
using that $\D(\oX)$ is a core for $\oX^{1/2}$ and $\bra u \oX \ket u = \norm{\oX^{1/2} u}^2$.
By linearity, we then get that for any $\orho \in \K_d$ such that $\Tr{\orho \oX} < \infty$,
\begin{equation}
        \Tr{\S_t(\orho)(\oX \wedge r)
        + C_2 \int_0^t \S_{s}(\orho)( \oX \wedge r)ds}
	\leq \Tr{\orho \oX} +C_1 t
	\label{eq:evol_regularized_bounds}
\end{equation}
and in particular $\Tr{\S_t(\orho) (\oX\wedge r) } \leq  \Tr{\orho \oX} +C_1 t$.
Taking the supremum over $r$, we also have
\begin{equation}
\Tr{\S_t(\orho) \oX} \leq \Tr{\orho \oX} +C_1 t < \infty.
\end{equation}

Let us now assume that $\Tr{\S_{t_0}(\orho_0) \oX} > {C_1}/{C_2}$ at some time $t_0\geq 0$;
then, for $r>0$ large enough, $\Tr{\S_{t_0}(\orho_0) (\oX\wedge r)} > {C_1}/{C_2}$.
Note that $\oX\wedge r$ is bounded and $\T_t$ is ultraweakly continuous,
so that $t \mapsto \Tr{\S_{t}(\orho_0) (\oX\wedge r)}$ is continuous and bounded.
Let $\bar{t}_r\in (t_0,+\infty]$ be the largest time such that
$\forall t\in [t_0,\bar{t}_r)$,
$\Tr{\S_t(\orho_0) (\oX\wedge r)} >{C_1}/{C_2}$. Note that $r \mapsto \bar t_r$ is increasing.
Then for any $t_1, t_2$ with $t_0\leq t_1\leq t_2 < \bar{t}_r$,
evaluating \cref{eq:evol_regularized_bounds}
at $t = t_2 - t_1$ and $\orho = \S_{t_1}(\orho_0)$,
we find
\begin{align}
	& \Tr{\S_{t_2}(\orho_0)(\oX\wedge r)}
		+ C_2 \int_{t_1}^{t_2} \Tr{\S_{s}(\orho_0)(\oX\wedge r)}ds
		\leq \Tr{\S_{t_1}(\orho_0)\oX} + C_1(t_2-t_1)\\
	\Rightarrow\quad &
	\Tr{\S_{t_2}(\orho_0)(\oX\wedge r)}
		\leq
		\Tr{\S_{t_1}(\orho_0)\oX}
		+ C_2 \int_{t_1}^{t_2} \left( \frac{C_1}{C_2} - \Tr{\S_{s}(\orho_0)(\oX\wedge r)} \right)ds \notag \\
		& \hspace{8.8em} \leq \Tr{\S_{t_1}(\orho_0)\oX}
\end{align}
and then, taking the limit $r\rightarrow +\infty$,
\begin{equation}
	\Tr{\S_{t_2}(\orho_0)\oX} \leq \Tr{\S_{t_1}(\orho_0)\oX}.
\end{equation}
The function $t\mapsto \Tr{\S_t(\orho_0)\oX}$ is thus decreasing on $[t_0, \sup_r \bar t_r)$.

This shows that for any $t\geq 0$, we have
\begin{equation}
	\Tr{\S_t(\orho_0) \oX} \leq \max\left\{\Tr{\orho_0 \oX}, \frac{C_1}{C_2}\right\}
		\leq \Tr{\orho_0 \left(\frac{C_1}{C_2}+\oX \right)}
\end{equation}
from which we see that
$\Tr{\S_t(\orho_0) \oX}$,
and thus also
$\Tr{\S_t(\orho_0) \oV}$,
is bounded uniformly with time, concluding the proof of \cref{prop-Vbounded}.

\subsubsection{From Proposition \ref{prop-Vbounded} to Hypothesis \ref{hyp-tightness}}
\label{sec-proof-coro-compactness}

Let us first consider
$\orho_0 \in \K^f\cap \K_d$
and fix $\epsilon>0$.
Using \cref{prop-Vbounded}, we have $\sup_t \Tr{\oV \orho_t}< C$
for some constant $C>0$ depending on $\orho_0$.
Following the end of
\cref{sec-compact_intuition_alpha0},
we decompose $\orho_t$ in the Fock basis as
\begin{align*}
\orho_t= \sum_{n_1,n_2,m_1,m_2 \in \N}
		\orho_{n_1,n_2,m_1,m_2}(t)
		\ket{n_1}\bra{n_2}\otimes \ket{m_1}\bra{m_2}
\end{align*}
so that, since $\oV$ is diagonal in the Fock basis:
\begin{align*}
    \Tr{\oV \orho_t}= \sum_{n,m\in\N}
	(\tfrac{n}{k}+m)^k \orho_{n,n,m,m}(t) \leq C.
\end{align*}
Hence, there exists $N>0$ such that
\begin{align*}
	\forall t\geq 0, \quad
	\sum_{\substack{n,m\in\N \\ \frac{n}{k}+m >  N}} \orho_{n,n,m,m}(t) \leq \epsilon,
\end{align*}
using the inequality
\begin{equation}
	\begin{aligned}
	\sum_{\substack{n,m\in\N \\ \frac{n}{k}+m >  N}} \orho_{n,n,m,m}(t)
		&= \sum_{\substack{n,m\in\N \\ \frac{n}{k}+m >  N}} \frac{\left(\tfrac nk + m\right)^k \orho_{n,n,m,m}(t)}
			{\left(\tfrac nk + m\right)^k} \\
		&\leq \frac{1}{N^k} \left( \sum_{\substack{n,m\in\N \\ \frac{n}{k}+m >  N}} \left(\tfrac nk + m\right)^k \orho_{n,n,m,m}(t) \right)\\
		&\leq \frac{1}{N^k} \Tr{\oV \orho_t} \leq \frac{C}{N^k}.
	\end{aligned}
\end{equation}
Defining the finite-dimensional subspace $E$ as
\begin{equation}
	E = \Span{\Ket{n}\otimes\Ket{m}\mid \frac{n}{k}+m \leq N}
\end{equation}
and denoting $\projE$ the associated orthogonal projector,
we thus have
\begin{align}
    \forall t\geq 0, \quad
	\Tr{\orho_t \projE}
	\geq 1-\epsilon
\end{align}
which concludes the proof for the case $\orho_0 \in \K^f\cap \K_d$.
The extension to an arbitrary initial state $\orho_0\in \K_d$ follows easily.
Indeed, by density of $\K^f\cap \K_d$ in $\K_d$,
there exists $\tilde \orho_0 \in \K^f\cap \K_d$
such that $\trnorm{\orho_0-\tilde \orho_0} \leq \frac{\epsilon}{2}$.
Applying the previous result to $\tilde \orho_0$,
there exists a finite-dimensional subspace $E$ such that
\begin{align*}
	\forall t\geq 0, \quad
    \Tr{\S_t (\tilde \orho_0) \projE} \geq 1-\frac{\epsilon}{2}.
\end{align*}
As $\S_t$ is a contraction semigroup on $\K^1$, we find
\begin{align*}
	\Tr{\orho_t \projE} &=
	\Tr{\S_t (\orho_0) \projE}
	=\Tr{\S_t (\tilde \orho_0) \projE}
		+ \Tr{\S_t (\orho_0 -\tilde \orho_0) \projE}\\
	& \geq 1-\frac{\epsilon}{2} - \trnorm{\S_t (\orho_0 -\tilde \orho_0)}
			\norm{\projE}_{\mathcal B(\H)}\\
	&\geq  1-\frac{\epsilon}{2}- \trnorm{\S_t (\orho_0 -\tilde \orho_0)}
	\geq  1-\frac{\epsilon}{2}- \trnorm{\orho_0 -\tilde \orho_0}\\
	&\geq 1-\epsilon.
\end{align*}

\subsection{Density result}
\label{sec-density}

The goal of this section is to prove that \cref{hyp-density} holds,
\emph{i.e.}, that the subspace
		\begin{equation}
		\E_\oL = \Span{ P(\oG^\dag,\ob^\dag) \Ket{v}\otimes \Ket{0}
			\mid
			P \in\C\langle X,Y\rangle,
			\ket v\in \ker{\oL}}
		\end{equation}
is dense in $\H$, where $\oG =  -i\oH - \frac\kappa2\ob^\dag\ob$.
\newline

This is rather straightforward when $\alpha=0$, and we will see
that the case $k=1$ with arbitrary $\alpha\in\R$ can then be deduced;
this is the subject of \cref{sec-alpha0k1}.
However, major difficulties arise for $k\geq 2$.
The main difference with the previous case is that, while
the operator $\oa-\alpha\Id$ is unitarily equivalent to $\oa$ for any $\alpha\in \R$,
the operators $\oa^k - \alpha^k\Id$ and $\oa^k$
are \emph{not} unitarily equivalent for $k\geq 2$,
so that the general case $\alpha\in\R$ can no longer be deduced from the specific case
$\alpha=0$;
its treatment will require the introduction of novel tools borrowed from complex analysis.
For the sake of pedagogy, we start in
\cref{sec-density-2}
with the case $k=2$,
before generalizing the proof to $k\geq 3$
in \cref{sec-proof-density-gen}.

\subsubsection{\texorpdfstring{Case $\alpha=0$ or $k=1$}{Case \unichar{"1D736}=0 or k=1}}
\label{sec-alpha0k1}
We start with the trivial remark that
\begin{equation}
	\C[\ob^\dag]\ket 0
		= \left\{ P(\ob^\dag) \ket 0 \; \mid \; P\in\C[X] \right\}
\end{equation}
is dense in $\H_b$,
which is a direct consequence of the definition of the Fock basis,
along with the relation $\ob^\dag \ket n = \sqrt{n+1} \ket{n+1}$
for all $n\in\N$.

It is natural to wonder if we can similarly find a family of non-commutative
polynomials $P_n\in\C\langle X,Y\rangle$
such that:
\begin{itemize}
	\item for all $n\in\N$,
		$P_n(\oG^\dag,\ob^\dag)$ is an operator on $\H_a$ only,
		in the sense $P_n(\oG^\dag,\ob^\dag) = \oO_n\otimes \Id_b$
		for some operator $\oO_n$ on $H_a$;
	\item the space
		\begin{equation}
			\E_\oL^a = \Span{ P_n(\oG^\dag,\ob^\dag) \Ket{v}
			\; \mid \;
			\ket v\in\ker\oL, n\in \mathbb N}
		\end{equation}
		is dense in $\H^a$.
\end{itemize}
If the answer to this question is positive, the density of $\E_\oL$
can be immediately deduced using the family of polynomials $P_{n,m}$
defined as $P_{n,m}(X,Y) = Y^m \,  P_n(X,Y)$.
In other words, we are wondering if the density of $\E_\oL$ in $\H = \H_a\otimes \H_b$
can be reduced to that of a given subspace $\E_\oL^a$ in $\H_a$ only.
\newline

Let us now assume that $\alpha=0$.
Using \cref{eq-kernel_cat_0} we know that
\begin{equation}
	\label{eq-kerL0}
	\ker\oL = \Span{\ket 0 , \ldots, \ket{k-1}}.
\end{equation}
Moreover, using the definition of $\oG$, we find
\begin{equation}
	\label{eq-commut_Ldag}
	-i \, \left[ \oG^\dag, \ob^\dag \right] - \tfrac{i\kappa}2 \ob^\dag
	= \left[ \oL \ob^\dag + \oL^\dag \ob, \ob^\dag \right]
		+ \tfrac{i\kappa}2 \left[ \ob^\dag \ob, \ob^\dag\right]
		- \tfrac{i\kappa}2 \ob^\dag
	= \oL^\dag
	= \oa^{\dag k}
\end{equation}
and, from the definition of Fock states,
\cref{eq-kerL0}
ensures that the space
\begin{equation}
	\left\{ (\oa^{\dag k})^n \ket v \; \mid \; \ket v \in\ker\oL, \, n\in\N \right\}
\end{equation}
is dense in $\H^a$.
In other words, the choice
\begin{equation}
	\label{eq:pn_polynomials}
	P_n(X,Y) = \left( -i \, (XY-YX) - \tfrac{i\kappa}2 Y\right)^n
\end{equation}
leads to $P_n(\oG^\dag, \ob^\dag) = \oL^{\dag n} = \oa^{\dag kn}$,
and we obtain that
\begin{equation}
	\label{eq-dense_Ldag0}
	\E_\oL^a 		= \Span{(\oL^\dag)^j \ket v
			\; \mid \;
			\ket v \in \ker\oL, \, j\in\N}
\end{equation}
is dense in $\H^a$,
which concludes the proof of \cref{hyp-density} in the case $\alpha=0$.
\newline

Let us now turn to the case $k=1$, that is $\oL = \oa - \alpha \Id$
for some $\alpha \in \R$.
We introduce the so-called Glauber displacement operator
\begin{equation}
	\oD_\alpha = e^{\alpha \oa^\dag - \alpha \oa}
\end{equation}
and remind the reader that $\oD_\alpha$ is unitary with $\oD_\alpha^\dag = \oD_{-\alpha}$,
and moreover satisfies the conjugation relation
\begin{equation}
	\oD_{-\alpha} \, \oa \, \oD_{\alpha} = \oa + \alpha\Id
\end{equation}
from which we deduce
\begin{equation}
	\oD_{-\alpha} \, \oL \, \oD_{\alpha}
	= \oD_{-\alpha} \,  \left( \oa - \alpha\Id \right) \, \oD_{\alpha}
	= \oa
\end{equation}
and
\begin{equation}
	\ker\oL = \oD_{\alpha} \, \ker\oa.
\end{equation}

This unitary transformation allows us to exploit the result obtained previously for $\alpha=0$.
Indeed, using the same polynomials $P_n$ (defined in \cref{eq:pn_polynomials}),
we have
\begin{equation}
	P_n(\oG^\dag, \ob^\dag) = \oL^{\dag n} = (\oa^\dag - \alpha\Id)^n
= \oD_\alpha \, \oa^{\dag n} \, \oD_{-\alpha}.
\end{equation}
The subspace $\E_\oL^a$ is then trivially dense as the image of a dense subspace
by the unitary operator $\oD_\alpha$:
\begin{equation}
	\begin{aligned}
	\E_\oL^a
		&=
		\Span{(\oL^\dag)^j \ket v
			\; \mid \;
			\ket v \in \ker\oL, \, j\in\N}\\
	&= \oD_\alpha \;
		\Span{ \oa^{\dag j} \ket 0
			\; \mid \;
			\, j\in\N}
	= \oD_\alpha \Span{ \ket j \; \mid \; j\in \N}.
	\end{aligned}
\end{equation}

This concludes the proof of \cref{hyp-density} in the case $k=1$.

\subsubsection{The Bargmann--Fock space}
\label{sec-bargmann-fock}

Let us remind the definition of the Bargmann--Fock space,
also known as Segal--Bargmann or sometimes simply Fock space;
we refer the reader to \cite[Chapter 2]{zhuAnalysisFockSpaces2012}
for a more complete introduction.
\begin{definition}
    The Hilbert space $\F^2$ is
	the set of holomorphic functions on $\C$ which belong to
	$L^2\left(\C,\frac{1}{\pi} e^{-|z|^2} dz\right)$,
	endowed with the inner product
    \begin{align*}
        \braket{f|g}_{\F^2}= \frac{1}{\pi}\int_{\C} \overline{f(z)}g(z) e^{- |z|^2} dz,
    \end{align*}
    where $dz$ denotes the Lebesgue measure on $\R^2$ identified with the complex plane.
\end{definition}
There is a correspondence between $\F^2$ and $\H_a$ thanks to the isometry
\begin{equation}
	\label{eq-iso_fock_space}
\begin{aligned}
    \left\{ \begin{array}{ccc}
        \H_a & \to &\F^2\\
	    \ket u = \sum_{n\in\N} u_n \ket n
		&\mapsto & \left( z \mapsto \sum_{n\in\N} \frac{u_n}{\sqrt{n!}} z^n \right)
    \end{array}.
    \right.
\end{aligned}
\end{equation}
Besides, the annihilation and creation operators on
$\H_a$
are respectively mapped to the derivation $\partial_z$ and the multiplication by $z$ on $\F^2$;
and a coherent state $\Ket{\beta}\in\H_a$
is represented by the function $z\mapsto e^{\beta z-\frac{|\beta|^2}{2}}$.

Finally, a core feature of $\F^2$ is that it is a reproducing kernel Hilbert space:
\begin{align}
	\label{eq-rkhs}
    \forall \beta\in\C, \forall f\in\F^2, \quad
		\braket{z\mapsto e^{\overline \beta z}|f}_{\F^2}=f(\beta).
\end{align}
In particular, taking scalar products with a coherent state $\ket \beta$ in $\H_a$
corresponds to an evaluation at $\overline \beta$ in $\F^2$.
Note that, using the fact that $\partial_z$ is the adjoint of the multiplication by $z$
and that $z\mapsto z^n e^{\overline\beta z}$ is in $\F^2$
for any integer $n$, we can extend
\cref{eq-rkhs}
to:
\begin{equation}
    \forall \beta\in\C, \, n\in\N, \, f\in\F^2, \quad
		\frac1\pi \int_\C
			\overline{ e^{\overline\beta z}} \, \partial_z^n f(z) e^{-|z|^2} dz
		=\partial_z^n f(\beta)
\end{equation}
where the left-hand side in the above equation is not a scalar product in $\F^2$
since $\partial_z^n f \notin \F^2$.
Exploiting this extended formula,
we will abuse notations and still write the left-hand side as
$\braket{z \mapsto e^{\overline\beta z} \, |\, \partial_z^n f}$.
\newline

In what follows, we will often identify elements of $\H_a$ and their
representation in $\F^2$.
Additionally, to alleviate notations, we will identify functions
with their evaluations, writing for instance $z^2$ instead of $z\mapsto z^2$
wherever $z$ is a free variable in context.

For a given function $f\in\F^2$, we will note $Z_f$ its set of zeros,
and more precisely $Z_f^k$ its set of zeros of order $k\geq 1$:
\begin{align}
	Z_f^k &= \left\{ z \in \C \; \mid \;
		\forall j \in \llbracket 0, k-1 \rrbracket, \, \partial_z^j f(z) =0;
		\quad \partial_z^k f(z) \neq 0\right\},\\
	Z_f &= \cup_{k\geq 1} Z_f^k = \left\{ z\in \C \; \mid \; f(z) = 0\right\}.
\end{align}
\newline

We will see that in the setting of engineered multi-photon
processes,
the density \cref{hyp-density} is related to the following question of
polynomial approximation in the Bergmann--Fock space $\F^2$,
sometimes called the \emph{Newman--Shapiro problem}:
given a function $f\in\F^2$,
we want to determine
whether the set of polynomial multiples of $f$
is dense
in the set of holomorphic multiples of $f$, \emph{i.e.}, whether we have:
\begin{equation}
	\label{eq-ns_density}
	\holofun f \cap \F^2 = \overline{\C[z] f \cap \F^2}
\end{equation}
where $\holofun$ denotes the set of all holomorphic functions
and the closure is for the $\F^2$ topology.
A recent result shows that \cref{eq-ns_density} generically does not hold,
and describes different classes of functions $f$ for which it does
\cite{belovNewmanShapiro2022}.
A specific case,
that will be useful in the following sections
and was already proven in the original papers introducing the problem
\cite{newmanCertainHilbert1966,newmanmimeographed},
states that \cref{eq-ns_density} holds whenever $f$ is an exponential polynomial.
In this case, we can further give a characterization of the orthogonal complement
of $\overline{\C[z] f \cap \F^2}$:
\begin{theorem}[\cite{newmanCertainHilbert1966}, Theorem 2; %
	see also \cite{belovNewmanShapiro2022}, Theorems 1.1 and 1.5]
	\label{thm-newman_shapiro}
	Let $f$ be an exponential polynomial, that is
	\begin{equation}
		f(z) = \sum_{0\leq k\leq d} P_k(z) e^{\lambda_k z}
	\end{equation}
	for some integer $d\in\N$ and some
	families of polynomials $P_k\in\C[X]$ and complex numbers $\lambda_k\in\C$;
	note that in particular $\C[z] f \subset \F^2$.
	Then:
	\begin{enumerate}
		\item
			\(
				\holofun f \cap \F^2 = \overline{\C[z] f},
			\)
		\item
			\(
				\mathcal M_f^\dagger = f^*(\partial_z),
			\)
			\\
			where $\mathcal M_f$ denotes the operator of multiplication by $f$
			defined on $\D(\mathcal M_f) = \{ g \in \F^2 \mid fg \in \F^2\}$,
			$\mathcal M_f^\dagger$ is its adjoint,
			and $f^*(\partial_z)$ is a formal notation for the adjoint of the restriction
			of $\mathcal M_f$ to finite sums of exponentials,
			that is $\Span{e^{\lambda z} \mid \lambda \in\C}$.
		\item \( \holofun f \cap \F^2 = \ker{\mathcal M_f^\dagger}^\perp\),
		\item
			\( \ker{\mathcal M_f^\dagger} \) is spanned by the exponential monomials
			(\emph{i.e.}, functions of the form $z^n e^{\lambda z}$) it contains.
			In particular, when all zeros of $f$ are simple, we obtain
			\begin{equation}
			\ker{\mathcal M_f^\dagger} =
				\overline{\Span{ e^{\overline{\lambda}z} \, \mid \, f(\lambda) =0}}.
			\end{equation}
			More generally, we have
			\begin{equation}
			\ker{\mathcal M_f^\dagger} =
				\overline{\Span{ z^j \, e^{\overline{\lambda}z}
					\, \mid \,
					\exists k>j, \lambda\in Z_f^k}}.
			\end{equation}
	\end{enumerate}
\end{theorem}

\subsubsection{Case $k=2$}
\label{sec-density-2}

Let us now consider the case $k=2$ and assume that $\alpha\neq 0$,
since if $\alpha=0$ the proof of the previous section applies.
Contrary to the case $k=1$, the operator $\oL=\oa^k - \alpha^k\Id$ is not unitarily
equivalent to $\oa^k$:
indeed, the subspace
\(\Span{(\oL^\dag)^j \ket v
			\; \mid \;
			\ket v \in \ker\oL, \, j\in\N}
\)
that appeared in \cref{eq-dense_Ldag0}
is no longer dense in $\H_a$,
whereas this property would be preserved by unitary equivalence.
However, we can explicitly characterize a complement of this subspace through
the following lemma:
\begin{lemma}
    \label{lem_non_stability_2legs}
	Let $\alpha\in\R\setminus\{0\}$, $k=2$ and $\oL = \oa^2 - \alpha^2\Id$.
	Define the following subspaces of $\H_a$:
	\begin{align}
		\label{eq-def_rl}
		\E_\oL^a &= 			\Span{(\oL^\dag)^j \ket{v} \mid j\in \N,\, v\in \ker{\oL}},\\
		\label{eq-def_rl_sharp}
		\E_\oL^\sharp &=
								\Span{(\oL^\dag)^j [\oL,\oL^\dag] \ket{v}
					\mid j\in \N,\, v\in \ker{\oL}}.
	\end{align}
	Then, $\E_\oL^a + \E_\oL^\sharp$ is dense in $\H_a$.
\end{lemma}
Comparing \cref{eq-def_rl,eq-def_rl_sharp}
with \cref{eq-dense_Ldag0} appearing earlier,
remark that in the case $k=1$, $\E_\oL^a$ was already dense,
whereas here, for $k=2$,
we have to add $\E_\oL^\sharp$ to span the whole space.
Additionally,
we saw in \cref{eq-commut_Ldag} that $\oL^\dag \in \C\langle \oG^\dag , \ob^\dag \rangle$
so that for all $n$, $\E_\oL^a \otimes \ket n \subset \E_\oL$.
On the other hand, the operator $[\oL, \oL^\dag]$
in the definition of $\E_\oL^\sharp$
is not directly a polynomial in $\oG^\dag$ and $\ob^\dag$,
but we have
\begin{equation}
	\label{eq-lldag_from_commut}
	-i \, \left[\oG^\dag, \, \oL^\dag \right] = \left[ \oL, \oL^\dag \right] \ob^\dag
\end{equation}
which is not an operator on $\H_a$ only due to the factor $\ob^\dag$.
As a result, we will have to justify how the density of $\E_\oL$ in $\H$
is deduced from
that of $\E_\oL^a+\E_\oL^\sharp$
in $\H_a$.

\begin{proof}[Proof of \cref{lem_non_stability_2legs}]
Let us first make precise how \cref{lem_non_stability_2legs} translates
in $\F^2$ through the isometry defined in \cref{eq-iso_fock_space}.
Using \cref{eq-kerL_orthobasis} and denoting with $\eqF $ the identification
of elements of $\H_a$
with their associated function in $\F^2$, we know that
\begin{equation}
	\ker\oL = \Span{ \ket\alpha + \ket{-\alpha}, \, \ket\alpha - \ket{-\alpha}}
	\eqF \Span{ \cosh(\alpha z), \, \sinh(\alpha z)}.
\end{equation}
Moreover, we have
\begin{align}
	\oL &= \oa^2 -\alpha^2\Id \eqF  \partial_z^2 - \alpha^2,\\
	\oL^\dag &= \oa^{\dag 2} -\alpha^{2}\Id \eqF z^2 - \alpha^{2},\\
	[\oL,\oL^\dag] &= [\oa^2,\oa^{\dag 2}] \eqF [\partial_z^2, z^2]
			= 4 z\, \partial_z + 2.
\end{align}
From the previous relations, we have
\begin{align}
	\E_\oL^a &= 		\Span{(\oL^\dag)^j \ket v \; \mid \; \ket v \in \ker\oL, \, j\in\N}\\
		&=\Span{(\oL^\dag + \alpha^{2}\Id)^j \ket v \; \mid \; \ket v \in \ker\oL, \, j\in\N}\\
				\label{eq-def_ela_holo}
		&\eqF
		\Span{
			 P(z^2) \, \cosh(\alpha z)
				+ Q(z^2) \, \sinh(\alpha z)
				\; \mid \; P,Q \in \C[X]
		},\\
	\\
	\E_\oL^\sharp &= 			\Span{(\oL^\dag)^j [\oL,\oL^\dag] \ket{v} \mid j\in \N,\, v\in \ker{\oL}}\\
		\label{eq-def_elsharp_holo}
		&\eqF
		\Span{
			P(z^2)
				\left(\cosh(\alpha z) + 2z\sinh(\alpha z)\right)
			+ Q(z^2)
				\left(\sinh(\alpha z) + 2z\cosh(\alpha z)\right)
			\; \mid \;
			P,Q \in \C[X]
		}.
\end{align}
Using \cref{eq-def_ela_holo,eq-def_elsharp_holo}
and regrouping terms,
we have
\begin{align}
	\label{eq-rewrite_target_space_in_F2}
	\E_\oL^a + \E_\oL^\sharp
	\eqF&
	\Span{P(z^2) \cosh(\alpha z) + Q(z^2) \,z \sinh(\alpha z)
		\mid P,Q \in\C[X] } \notag\\
	&\oplus
	\Span{P(z^2) \sinh(\alpha z) + Q(z^2)\, z \cosh(\alpha z)
		\mid P,Q \in\C[X] }
\end{align}
	and \cref{lem_non_stability_2legs}
	is equivalent to $( \E_\oL^a + \E_\oL^\sharp)^\perp = \{0\}$.

	Let us consider $\ket\psi \in ( \E_\oL^a + \E_\oL^\sharp)^\perp$
	and define $\psi$ the corresponding element of $\F^2$, that is $\ket\psi \eqF \psi$.
	Using the fact that $e^{\alpha z} = \cosh(\alpha z) + \sinh(\alpha z)$
	and that any polynomial $P\in\C[X]$ can be decomposed as a sum
	$P(X) = P_1(X^2) + X P_2(X^2)$ with $P_1,P_2\in\C[X]$,
	we see from \cref{eq-rewrite_target_space_in_F2}
	that $\ket \psi \in ( \E_\oL^a + \E_\oL^\sharp)^\perp$ implies:
	\begin{equation}
		\psi \perp \C[z] e^{\alpha z}.
	\end{equation}
	Using the Newman-Shapiro result in \cref{thm-newman_shapiro}
	and the fact that $e^{\alpha z}$ does not cancel,
	we immediately deduce from the previous equation that
	\begin{equation}
		\psi \in \ker{M_{e^{\alpha z}}^\dagger}
			= \{0\}.
	\end{equation}
\end{proof}

The above proof sheds a new light on the lack of density of $\E_\oL^a$ when $k=2$:
identified with its representation in the space $\F^2$,
$\E_\oL^a$ does not contain all polynomial multiples of $e^{\alpha z}$,
but only those associated with even polynomials.
The new elements introduced by $\E_\oL^\sharp$,
which add odd polynomials, are then required to reconstruct all of $\C[z]$.

\paragraph{From \cref{lem_non_stability_2legs} to \cref{hyp-density}}
We now want to prove that
\begin{align*}
    \E_\oL = \Span{ P(\oG^\dag,\ob^\dag) \Ket{v}\otimes \Ket{0}
		\; \mid \; P\in\C\langle X,Y\rangle,\, \ket v\in \ker{\oL}}
\end{align*}
is dense in $\H$,
with $\oG = -i \oH - \frac\kappa2 \ob^\dag\ob$.
As $\H_\oL = \ker{\oL}\otimes \Ket{0}\subset \H^{n,n}$ for every $n\in \N$,
we have $\E_\oL \subset \D(\oG^\infty)\cap \D(\ob^\infty)$.
Hence, we implicitly restrict
all the operators to the dense subset $\D(\oG^\infty)\cap \D(\ob^\infty)$
in the following computations.

Let us consider $\ket\psi \in \E_\oL^\perp$; we want to prove that, necessarily, $\ket\psi =0$.
Firstly, recall from
\cref{eq-commut_Ldag,eq-lldag_from_commut} that
\begin{align}
    \label{eq-bracket_L_dag-2leg}
	&-i\, [\oG^\dag, \ob^\dag] -\frac{i\kappa}2 \ob^\dag
	=\oL^\dag, \\
    \label{eq-bracket_LL_dag-2leg}
	&-i [\oG^\dag,\oL^\dag]
	=  [\oL,\oL^\dag]\ob^\dag.
\end{align}

    Since $\ket\psi \in\E_\oL^\perp$,
    \cref{eq-bracket_L_dag-2leg,eq-bracket_LL_dag-2leg}
    imply that for any $P\in \C[X]$, $\ket v \in \ker{\oL}$ and $n\in \N$:
    \begin{align}
        \label{eq-ortho_psi1-2leg}
        \Bra{\psi} P(\oL^\dag) (\ob^\dag)^{n} \ket v\otimes \Ket{0}=0, \\
        \label{eq-ortho_psi2-2leg}
        \Bra{\psi} P(\oL^\dag) [\oL,\oL^\dag] (\ob^\dag)^{n+1} \ket v\otimes \Ket{0}=0.
    \end{align}
    Let us now decompose $\ket \psi$ along the eigenspaces of $\ob^\dag \ob$:
    \begin{align}
	    \label{eq-decomposing_psi_bdagb}
	    \Ket{\psi}=\sum_{m=0}^\infty \Ket{\psi_m}\otimes\ket{m} ,\qquad \ket{\psi_m}\in \H_a.
    \end{align}
    For any $m\geq1$,
    noting that $(\ob^\dag)^m \ket 0 = {\sqrt{m!}}\ket m$ and
    injecting \cref{eq-decomposing_psi_bdagb} into \cref{eq-ortho_psi1-2leg,eq-ortho_psi2-2leg},
    with $n=m$ and $n=m-1$ respectively,
    we obtain that, for all $\ket v \in \ker{\oL}$ and $P\in\C[X]$, we have
    \begin{align}
	    \bra{\psi_m} P(\oL^\dag) \ket v = 0,\\
	    \bra{\psi_m} P(\oL^\dag) [\oL,\oL^\dag] \ket v = 0
    \end{align}
    from which \cref{lem_non_stability_2legs} allows deducing that $\ket{\psi_m}\in (\E_\oL^a)^\perp \cap (\E_\oL^\sharp)^\perp = \{0\}$.
    Hence, \cref{eq-decomposing_psi_bdagb} boils down to $\Ket{\psi}=\Ket{\psi_0}\otimes\ket{0}$.
    Using \cref{eq-ortho_psi1-2leg} with $n=0$, we obtain
    \begin{equation}
\label{eq-psi0_ortho_Ela}
	    \ket{\psi_0} \perp \E_\oL^a.
    \end{equation}
    Moreover, we have the following equalities on linear forms,
    which are well-defined on $\ker\oL$ since it is included in
    $\D(\oG^{\infty})\cap \D(\ob^{\infty})$:
    \begin{align}
	    \bra \psi \oG^\dag &=
		\left( \bra{\psi_0}\otimes \bra 0 \right)
		\left( i (\oL \ob^\dag + \oL^\dag \ob) - \frac\kappa2 \ob^\dag \ob\right)\\
	    &=i \left(\bra{\psi_0} \oL^\dag \right)\otimes \bra 1,\\
	    \bra\psi \oG^{\dag 2} &=
	    - \sqrt2 \left( \bra{\psi_0} \oL^{\dag 2} \right) \otimes \bra 2
	    - \left( \bra{\psi_0} \oL^\dag \oL\right) \otimes \bra 0
	    - \frac{i\kappa}2 \left( \bra{\psi_0} \oL^\dag\right)\otimes \bra1.
    \end{align}
    Hence, for any $\ket v\in\ker\oL$ and $P\in\C[X]$,
    using the last equation and noting that
    \begin{equation}
	    \oG^{\dag 2} P\left( -i [\oG^\dag,\ob^\dag] - \frac{i\kappa}2 \ob^\dag\right)
	= \oG^{\dag 2} P(\oL^\dag) \in \C\langle \oG^\dag, \ob^\dag \rangle,
    \end{equation}
	we obtain
    \begin{align}
	    \label{eq-psi0_ortho_H2}
	    0 = \bra \psi \oG^{\dag 2} \, P(\oL^\dag) \left(\ket v\otimes \ket 0\right)
	= \bra{\psi_0} \oL^\dag \oL \, P(\oL^\dag)\ket v.
    \end{align}
    Denoting $\psi_0$ the function in $\F^2$ associated to $\ket{\psi_0}$,
    that is $\ket{\psi_0} \eqF \psi_0$,
    \cref{eq-psi0_ortho_Ela}
    translates to
    \begin{align}
	    \label{eq-psi0fun_ortho_ela}
	    \psi_0 \perp \Span{ P(z^2) \cosh(\alpha z) + Q(z^2) \sinh(\alpha z) \; \mid \; P,Q\in\C[X]}
    \end{align}
    while \cref{eq-psi0_ortho_H2}
	    translates to
	\begin{align}
		\label{eq-psi0fun_ortho_ldagl}
		\psi_0 \perp
		\mathrm{Span}\big\{& (z^{2}-\alpha^{2})(\partial_z^2-\alpha^2) P(z^2) \cosh(\alpha z) \notag\\
		& \quad + (z^{2}-\alpha^{2})(\partial_z^2-\alpha^2) Q(z^2) \sinh(\alpha z)
	    \mid P,Q \in \C[X]\big\}.
    \end{align}
    For any $P\in\C[X]$, we have the following operator relationship between
    $\partial_z^2$ and the multiplication by $P(z^2)$:
    \begin{align}
	    \partial_z^2 \, P(z^2)
		&= \partial_z \left( \partial_z P(z^2) \right)\\
		&= \partial_z \left( 2z\, P'(z^2) + P(z^2) \, \partial_z \right)\\
		&= 2 P'(z^2)
			+ 4z^2P''(z^2)
			+ 4\, P'(z^2) \,z \partial_z
			+ P(z^2) \, \partial_z^2
    \end{align}
    where $P'$ and $P''$ denote the first and second polynomial derivatives of $P$ in $\C[X]$,
    so that
    \begin{equation}
    \begin{aligned}
	    \label{eq-commut_poly_dz}
	    (z^{2}-\alpha^{2})(\partial_z^2-\alpha^2) P(z^2)
	    &= P(z^2) \, (z^{2}-\alpha^{2})(\partial_z^2-\alpha^2)\\
	    &\quad + \left( 2 P'(z^2) + 4z^2P''(z^2) \right) (z^{2}-\alpha^{2})\\
	    &\quad + 4 P'(z^2) \, z \, (z^2-\alpha^{2}) \,\partial_z.
    \end{aligned}
    \end{equation}
    Combining the relation $(\partial_z^2-\alpha^2)\cosh(\alpha z) = (\partial_z^2-\alpha^2)\sinh(\alpha z)=0$
    with
    \cref{eq-psi0fun_ortho_ldagl,eq-commut_poly_dz},
    and noting that any polynomial can be obtained as the derivative of another polynomial
    and that
    $\partial_z \cosh(\alpha z) = \alpha \sinh(\alpha z)$,
    $\partial_z \sinh(\alpha z) = \alpha \cosh(\alpha z)$,
    we obtain that for any $P,Q\in\C[X]$:
    \begin{align}
	    \label{eq-psi0_ortho_part1}
	    \psi_0 &\perp P(z^2) \, z \, (z^2-\alpha^{2}) \, \sinh(\alpha z),\\
	    \label{eq-psi0_ortho_part2}
	    \psi_0 &\perp Q(z^2) \, z \, (z^2-\alpha^{2}) \, \cosh(\alpha z).
    \end{align}
    From \cref{eq-psi0fun_ortho_ela}, we also obtain for any $P,Q\in\C[X]$:
    \begin{align}
	    \label{eq-psi0deriv_ortho_ela}
	    \psi_0 &\perp
	    P(z^2) (z^2-\alpha^{2}) \, \cosh(\alpha z),\\
	    \label{eq-psi0_ortho_part4}
	    \psi_0 &\perp
			Q(z^2) (z^2-\alpha^{2}) \, \sinh(\alpha z).
    \end{align}
In particular, since
$e^{\alpha z} = \cosh(\alpha z) + \sinh(\alpha z)$
and
one can decompose any polynomial $P\in\C[X]$
as $P(X) = P_1(X^2) + XP_2(X^2)$ with $P_1,P_2\in\C[X]$,
\cref{eq-psi0_ortho_part1,eq-psi0_ortho_part2,eq-psi0deriv_ortho_ela,eq-psi0_ortho_part4}
yield
\begin{equation}
	\psi_0 \perp \C[z] (z^2-\alpha^{2}) e^{\alpha z}.
\end{equation}
Applying the Newman-Shapiro result in \cref{thm-newman_shapiro}
to the function $(z^2-\alpha^{2}) e^{\alpha z}$,
which is an exponential polynomial with two simple zeros in $\pm\alpha$,
we obtain
\begin{equation}
	\psi_0 \in
	\overline{\Span{e^{\overline\lambda z} \mid (\lambda^2 - \alpha^{2}) \, e^{\alpha \lambda} =0}}
	= \Span{e^{\alpha z}, e^{-\alpha z}}
	= \Span{\cosh(\alpha z), \sinh(\alpha z)}
\end{equation}
    and finally, using the orthogonality relation in
    \cref{eq-psi0fun_ortho_ela},
    \begin{equation}
	    \psi_0 = 0.
    \end{equation}

    This concludes the proof of \cref{hyp-density} and thus \cref{th-convergence_cat_qubit} in the case $k=2$.

\subsubsection{\texorpdfstring{Case $k\geq 3$}{Case k\unichar{"2265}3}}
\label{sec-proof-density-gen}
Apart from technicalities,
the generalization of \cref{sec-density-2} to the case $k>2$
follows the same general structure:
we first obtain an explicit characterization
of a complement of $\E_\oL^a$ in $\H_a$
through a density result in $\H_a$,
and then explain how this density result in $\H_a$
entails the density result in $\H = \H_a \otimes \H_b$ required in \cref{hyp-density}.

\begin{lemma}
    \label{lem_non_stability_klegs}
	Let $\alpha\in\R\setminus\{0\}$, $k\geq 3$ and $\oL = \oa^k - \alpha^k\Id$.
	Define the following subspaces of $\H_a$:
	\begin{align}
		\label{eq-def_rl_kcat}
		\E_\oL^a &= 			\Span{(\oL^\dag)^j \ket{v} \mid j\in \N,\, v\in \ker{\oL}},\\
		\label{eq-def_rl_sharp_kcat}
		\E_\oL^\sharp &=
								\Span{(\oL^\dag)^j [\oL,\oL^\dag]^{(s)} \ket{v}
					\mid j\in \N, \, 1\leq s\leq k-1, \, v\in \ker{\oL}}
	\end{align}
	where $[A,B]^{(s)}$ denotes the $s$-th iterated right commutator of $A$ with $B$,
	that is
	\begin{align}
		[A,B]^{(1)} &= [A,B],\\
		[A,B]^{(2)} &= [[A,B],B],\\
		&\textrm{etc.}\notag
	\end{align}
	Then, $\E_\oL^a + \E_\oL^\sharp$ is dense in $\H_a$.
\end{lemma}
Comparing \cref{lem_non_stability_klegs}
with \cref{lem_non_stability_2legs} earlier,
we see that the only difference is that we now need up to $k-1$ iterated brackets
in the definition of $\E_\oL^\sharp$ in
\cref{eq-def_rl_sharp_kcat}.
This can be intuitively understood as such:
recall that
for $k=2$, seen in the Bargmann--Fock space,
$\E_\oL^a$ contained all polynomial multiples of $e^{\alpha z}$
corresponding to even polynomials $P(z^2)$,
and $E_\oL^\sharp$ was needed to introduced odd terms of the form $z P(z^2)$
and thus reach all of $\C[z]$.
We then expect that in the generic case,
$\E_\oL^a$ will contain all polynomial multiples of $e^{\alpha z}$
corresponding to polynomials of the form $P(z^k)$,
and the $k-1$ iterated brackets above will introduce
the intermediary powers $z P(z^k), z^2 P(z^k), \ldots, z^{k-1}P(z^k)$
needed to reach all of $\C[z]$.

\begin{proof}[Proof of \cref{lem_non_stability_klegs}]
Let us first make precise how \cref{lem_non_stability_klegs} translates
in $\F^2$ through the isometry defined in \cref{eq-iso_fock_space}.
Using \cref{eq-kernel_cat} and denoting with $\eqF $ the identification
of elements of $\H_a$
with their associated function in $\F^2$, we know that
\begin{equation}
	\ker\oL = \Span{ \ket{\omega^r \alpha}
		\; \mid \; 0\leq r \leq k-1
		}
	\eqF \Span{ e^{\omega^r \alpha z} \; \mid \; 0\leq r \leq k-1}
\end{equation}
	with $\omega = e^{\frac{2i\pi}k}$ a primitive $k$-th root of unity.
Moreover, we have
\begin{align}
	\oL &= \oa^k -\alpha^k\Id \eqF  \partial_z^k - \alpha^k,\\
	\oL^\dag &= \oa^{\dag k} -\alpha^{k}\Id \eqF z^k - \alpha^{k},\\
	[\oL,\oL^\dag] &= [\oa^k,\oa^{\dag k}] \eqF [\partial_z^k, z^k].
\end{align}

Let us consider $\ket\psi \perp \E_\oL^a + \E_\oL^\sharp$; we want to prove that necessarily
$\ket\psi=0$.
Denoting $\psi$ the corresponding element of $\F^2$, defined through $\ket\psi \eqF \psi$,
we see that $\ket\psi \perp \E_\oL^a + \E_\oL^\sharp$
	implies that, for any $P\in\C[X]$ and $s\geq 1$, we have:
	\begin{align}
		\label{eq-psi_perp_zk}
		\psi &\perp P(z^k) \, e^{\alpha z},\\
		\label{eq-psi_perp_zk_commut}
		\psi &\perp P(z^k) \, \left[\partial_z^k,z^k\right]^{(s)} \, e^{\alpha z}
	\end{align}
where,
	with a slight abuse of notations,
	$z$ denotes the operator of multiplication by $z$ in the above formulae.
	Using Leibniz derivation formula, we obtain
	\begin{align}
		\left[\partial_z^k, \, z^k\right]
		&= \partial_z^k \, z^k - z^k \, \partial_z^k\\
		&= \sum_{r=0}^{k-1} \binom{k}{r} \,  \partial_z^{k-r}(z^k) \, \partial_z^{r}\\
		&= \sum_{r=0}^{k-1} \binom{k}{r} \, \frac{k!}{r!} \, z^{r} \, \partial_z^{r}\\
		\label{eq-commut_zk_dzk}
		&= k^2 \, z^{k-1} \, \partial_z^{k-1} +
			\sum_{r=0}^{k-2} \binom{k}{r} \, \frac{k!}{r!} \, z^{r} \, \partial_z^{r}.
	\end{align}
	Similarly, for any $j\in \llbracket 1, k \rrbracket$:
	\begin{align}
		\left[z^j \partial_z^j, \, z^k\right]
		&= z^j \left[\partial_z^j, \, z^k\right]\\
		&= z^j \left(
			\sum_{r=0}^{j-1} \binom{j}{r} \, \partial_z^{j-r}(z^k) \, \partial_z^r
			\right)\\
		&= z^j \left(
			\sum_{r=0}^{j-1} \binom{j}{r} \, \frac{k!}{(k-j+r)!}\, z^{k-j+r} \, \partial_z^r
			\right)\\
		&= z^k \left(
			\sum_{r=0}^{j-1} \binom{j}{r} \, \frac{k!}{(k-j+r)!}\, z^{r} \, \partial_z^r
			\right)\\
		\label{eq-commut_zk_dzk_iterated}
		&= z^k \left( kj \, z^{j-1}\, \partial_z^{j-1} +
			\sum_{r=0}^{j-2} \binom{j}{r} \, \frac{k!}{(k-j+r)!}\, z^{r} \, \partial_z^r
			\right).
	\end{align}
	Combining \cref{eq-commut_zk_dzk,eq-commut_zk_dzk_iterated},
	we see that the family of differential operators
	$\left( \left[ \partial_z^k, \, z^k\right]^{(s)} \right)_{1\leq s\leq k-1}$
	has a triangular expansion in the basis $\left( z^j \, \partial_z^j \right)_{0\leq j\leq k-1}$
	up to global multiplicative factors that are powers of $z^k$:
	for any $s\in\llbracket 1,k-1\rrbracket$,
	there exists positive coefficients
	$(c_{s,j})_{0\leq j\leq k-s}$ with $c_{s,k-s} \neq 0$ such that
	\begin{equation}
		\label{eq-diffop_triang}
		\left[ \partial_z^k, \, z^k\right]^{(s)}
		= z^{k(s-1)} \left(
				c_{s,k-s} \, z^{k-s} \, \partial_z^{k-s}
				+ \sum_{r=0}^{k-s-1}
					c_{s,r} z^r \partial_z^r.
			\right)
	\end{equation}
	Thanks to this triangular structure, and noting that
	$\partial_z^j e^{\alpha z} = \alpha^j e^{\alpha z}$ for any $j\geq 0$,
	\cref{eq-psi_perp_zk,eq-psi_perp_zk_commut}
	imply that for any $P\in\C[X]$ and $s\in \llbracket 1,k-1\rrbracket $:
	\begin{align}
		\psi &\perp P(z^k) \, e^{\alpha z},\\
		\psi &\perp P(z^k) \, z^{k(s-1)}\, z^s e^{\alpha z}
	\end{align}
	and thus in particular
	\begin{equation}
		\label{eq-ortho_poly_with_prefactor}
		\psi \perp \C[z] \, z^{k(k-2)} \, e^{\alpha z}.
	\end{equation}

	Applying the Newman-Shapiro result in \cref{thm-newman_shapiro}
	to the function $ z^{k(k-2)} \, e^{\alpha z}$,
	which is an exponential polynomial with a unique zero
	of order $k(k-2)$ at the origin,
	we can deduce from \cref{eq-ortho_poly_with_prefactor} that
	\begin{equation}
		\psi \in \ker{M_{ z^{k(k-2)} \, e^{\alpha z}}^\dagger}
			= \C_{k(k-2)-1}[z],
	\end{equation}
	\emph{i.e.}, $\psi$ is a polynomial of order at most $k(k-2)-1$.

	Let us now exploit the initial assumption that
	$\ket\psi \in(\E_\oL^a)^\perp$,
	which implies that for any $r\in\llbracket 0, k-1 \rrbracket$,
	we have
	\begin{equation}
		\psi \perp P(z^k) \, e^{\omega^r \alpha z}, \quad \omega = e^{\frac{2i\pi}k}
	\end{equation}
	and thus in particular
	\begin{equation}
		0 = \braket{ \psi \; \mid \; z^{k(k-3)} \, e^{\omega^r \alpha z}}_{\F^2}
		= \braket{\partial_z^{k(k-3)} \psi \; \mid \; e^{\omega^r \alpha z}}_{\F^2}
		= \overline{ \partial_z^{k(k-3)} \psi( \overline{\omega^r \alpha}) }.
	\end{equation}
	The function $\partial_z^{k(k-3)} \psi$, which is a polynomial of order at most
	$k-1$, cancels on the $k$ distinct values $(\overline{\omega^r \alpha})_{0\leq r\leq k-1}$ and is thus
	identically null,
	so that $\psi$ is actually a polynomial of order at most
	$k(k-3)-1$.
	Iterating this argument
	with all derivatives $\partial_z^{k(k-j)}\psi$
	for $3 \leq j \leq k$,
	we finally conclude that $\psi=0$.

\end{proof}

\paragraph{From \cref{lem_non_stability_klegs} to \cref{hyp-density}}
We now want to prove that
\begin{align*}
    \E_\oL = \Span{ P(\oG^\dag,\ob^\dag) \Ket{v}\otimes \Ket{0}
		\; \mid \; P\in\C\langle X,Y\rangle,\, \ket v\in \ker{\oL}}
\end{align*}
is dense in $\H$,
with $\oG = -i \oH - \frac\kappa2 \ob^\dag\ob$.
As $\H_\oL = \ker{\oL}\otimes \Ket{0}\subset \H^{n,n}$ for every $n\in \N$,
we have $\E_\oL \subset \D(\oG^\infty)\cap \D(\ob^\infty)$.
Hence, we implicitly restrict
all the operators to the dense subset $\D(\oG^\infty)\cap \D(\ob^\infty)$
in the following computations.

Let us consider $\ket\psi \in \E_\oL^\perp$; we want to prove that, necessarily, $\ket\psi =0$.
Firstly, recall from
\cref{eq-commut_Ldag,eq-lldag_from_commut} that
\begin{align}
    \label{eq-bracket_L_dag-kleg}
	&-i\, [\oG^\dag, \ob^\dag] -\frac{i\kappa}2 \ob^\dag
	=\oL^\dag, \\
    \label{eq-bracket_LL_dag-kleg}
	&-i [\oG^\dag,\oL^\dag]
	=  [\oL,\oL^\dag]\ob^\dag
\end{align}
and, iterating from the last equation, for all $s\geq1$:
\begin{equation}
	-i [\oG^\dag,\oL^\dag]^{(s)}
	=  [\oL,\oL^\dag]^{(s)}\ob^\dag.
\end{equation}

    Since $\ket\psi \in\E_\oL^\perp$,
    \cref{eq-bracket_L_dag-kleg,eq-bracket_LL_dag-kleg}
    imply that for any $P\in \C[X]$, $\ket v \in \ker{\oL}$, $n\in \N$ and $s\geq 1$:
    \begin{align}
        \label{eq-ortho_psi1-kleg}
        \Bra{\psi} P(\oL^\dag) (\ob^\dag)^{n} \ket v\otimes \Ket{0}=0, \\
        \label{eq-ortho_psi2-kleg}
	    \Bra{\psi} P(\oL^\dag) [\oL,\oL^\dag]^{(s)} (\ob^\dag)^{n+1} \ket v\otimes \Ket{0}=0.
    \end{align}
    Let us now decompose $\ket \psi$ along the eigenspaces of $\ob^\dag \ob$:
    \begin{align}
	    \label{eq-decomposing_psi_bdagb_kleg}
	    \Ket{\psi}=\sum_{m=0}^\infty \Ket{\psi_m}\otimes\ket{m} ,\qquad \ket{\psi_m}\in \H_a.
    \end{align}
    For any $m\geq1$,
    noting that $(\ob^\dag)^m \ket 0 = \sqrt{m!}\ket m$ and
    injecting \cref{eq-decomposing_psi_bdagb_kleg} into \cref{eq-ortho_psi1-kleg,eq-ortho_psi2-kleg},
    with $n=m$ and $n=m-1$ respectively,
    we obtain that, for all $\ket v \in \ker{\oL}$,
    $s\geq 1$ and $P\in\C[X]$, we have
    \begin{align}
	    \bra{\psi_m} P(\oL^\dag) \ket v = 0,\\
	    \bra{\psi_m} P(\oL^\dag) [\oL,\oL^\dag]^{(s)} \ket v = 0
    \end{align}
    from which \cref{lem_non_stability_klegs} allows deducing that $\ket{\psi_m}\in (\E_\oL^a)^\perp \cap (\E_\oL^\sharp)^\perp = \{0\}$.
    Hence, \cref{eq-decomposing_psi_bdagb_kleg}
    boils down to $\Ket{\psi}=\Ket{\psi_0}\otimes\ket{0}$.
    Using \cref{eq-ortho_psi1-kleg} with $n=0$, we obtain
    \begin{equation}
\label{eq-psi0_ortho_Ela_kleg}
	    \ket{\psi_0} \perp \E_\oL^a.
    \end{equation}
    Moreover, for any $Q\in\C[X]$,
    we have the following equality on linear forms,
    which are well-defined on $\ker\oL$ since it is included in
    $\D(\oG^{\infty})\cap \D(\ob^{\infty})$:
    \begin{align}
	    \bra\psi Q(\oL^\dag) \oG^{\dag 2} &=
	    - \sqrt2 \left( \bra{\psi_0} Q(\oL^\dag) \oL^{\dag 2} \right) \otimes \bra 2\\
	    &\quad - \left( \bra{\psi_0} Q(\oL^\dag) \oL^\dag \oL\right) \otimes \bra 0\\
	    &\quad - \frac{i\kappa}2 \left( \bra{\psi_0} Q(\oL^\dag) \oL^\dag\right)\otimes \bra1.
    \end{align}
    Hence, for any $\ket v\in\ker\oL$ and $Q,P\in\C[X]$,
    using the last equation and noting that
    \begin{equation}
	    Q\left( -i [\oG^\dag,\ob^\dag] - \frac{i\kappa}2 \ob^\dag\right) \,
	    \oG^{\dag 2} \,
	    P\left( -i [\oG^\dag,\ob^\dag] - \frac{i\kappa}2 \ob^\dag\right)
	    = Q(\oL^\dag) \, \oG^{\dag 2} \, P(\oL^\dag) \in \C\langle \oG^\dag, \ob^\dag \rangle,
    \end{equation}
	we obtain
    \begin{align}
	    \label{eq-psi0_ortho_H2_kleg}
	    0 = \bra \psi Q(\oL^\dag) \, \oG^{\dag 2} \, P(\oL^\dag) \left(\ket v\otimes \ket 0\right)
	    = \bra{\psi_0} Q(\oL^\dag) \, \oL^\dag \oL \, P(\oL^\dag)\ket v.
    \end{align}
    In particular, for any $P\in\C[X]$ and $s\geq1$, we have
    \begin{align}
	    \label{eq-kleg_psi0_ortho}
	    \bra{\psi_0} \oL^\dag P(\oL^\dag) \ket v = 0, \\
	    \label{eq-kleg_psi0_ortho_deriv}
	    \bra{\psi_0} \oL^\dag P(\oL^\dag) [\oL, \oL^\dag]^{(s)} \ket v &=0.
    \end{align}
    The first line is a direct consequence of
    \cref{eq-psi0_ortho_Ela_kleg}.
    The second line is a consequence of
    \cref{eq-psi0_ortho_H2_kleg}
    since $\oL^\dag P(\oL^\dag) [\oL, \oL^\dag]^{(s)}$
    can be decomposed as a sum of polynomial terms in $\oL,\oL^\dag$ of the form
    $\tilde Q(\oL^\dag) \, \oL^\dag \oL \, \tilde P(\oL^\dag)$
    for some $\tilde Q, \tilde P \in \C[X]$.

    Let us introduce the function $\psi_0\in\F^2$ associated to the state $\ket{\psi_0}$,
    defined through $\ket{\psi_0}\eqF \psi_0$.
\cref{eq-kleg_psi0_ortho,eq-kleg_psi0_ortho_deriv} imply in particular that,
for any $P\in\C[X]$ and $s\in\llbracket 1,k-1 \rrbracket$,
we have
\begin{align}
	\psi_0 &\perp (z^k - \alpha^{k}) \, P(z^k) e^{\alpha z},\\
	\psi_0 &\perp (z^k - \alpha^{k}) \, P(z^k) \, \left[ \partial_z^k, \, z^k\right]^{(s)} \,
			e^{\alpha z}.\\
\end{align}
Re-using the triangular structure of the differential operators
$( \left[ \partial_z^k, \, z^k \right]^{(s)} )_{1\leq s\leq k-1}$
obtained in \cref{eq-diffop_triang},
the previous relations imply
\begin{align}
	\psi_0 &\perp P(z^k) \, (z^k - \alpha^{k}) \, e^{\alpha z},\\
	\psi_0 &\perp z^{k(s-1)} z^s \, P(z^k) \, (z^k - \alpha^{k}) \, e^{\alpha z},
	\quad \forall 1\leq s \leq k-1\\
\end{align}
from which we can finally deduce
\begin{equation}
	\label{eq-psi0_newman_shapiro_ortho_kleg}
	\psi_0 \perp \C[z] \, z^{k(k-2)} \, (z^k - \alpha^{k}) \, e^{\alpha z}.
\end{equation}
	Applying the Newman-Shapiro result in \cref{thm-newman_shapiro}
	to the function $ z^{k(k-2)} \, (z^k - \alpha^{k}) \, e^{\alpha z}$,
	which is an exponential polynomial with a zero
	of order $k(k-2)$ at the origin
	and $k$ simple zeros at $\omega^r \alpha$
	for $0\leq r \leq k-1$ and $\omega = e^{\frac{2i\pi}k}$,
	we can deduce from \cref{eq-psi0_newman_shapiro_ortho_kleg} that
	\begin{equation}
		\psi_0(z) = P(z) + \sum_{r=0}^{k-1} c_r e^{\omega^r \alpha z}
	\end{equation}
	for some coefficients $c_r\in\C$ and some polynomial $P$
	of order at most $k(k-2)-1$.
	We can now prove separately that $P$ and the $c_r$ coefficients are null.
	On the one hand, note that
	$\partial_z^{k(k-2)} P = 0$ while for any $r$,
	$\partial_z^{k(k-2)} e^{\omega^r \alpha z}
		= \omega^{k(k-2) r} \alpha^{k(k-2)} e^{\omega^r \alpha z}
		= \alpha^{k(k-2)} e^{\omega^r \alpha z}$
	so that
		\begin{equation}
			\label{eq-kleg_high_derivative_in_kerL}
			\partial_z^{k(k-2)} \psi_0 =
				\alpha^{k(k-2)} \sum_{r=0}^{k-1} c_r e^{\omega^r \alpha z}.
		\end{equation}
	On the other hand
	\cref{eq-psi0_ortho_Ela_kleg} implies:
	\begin{equation}
		\psi_0 \perp \Span{ P(z^k) e^{\omega^r \alpha z},
				\; \mid \; P\in\C[X], 0\leq r\leq k-1}
	\end{equation}
	so that for any $r\in\llbracket 0, k-1 \rrbracket$:
	\begin{equation}
		0 = \braket{ \psi_0 \, | \, z^{k(k-2)} e^{\omega^r \alpha z}}
			= \braket{ \partial_z^{k(k-2)} \psi_0 \, | \, e^{\omega^r \alpha z}},
	\end{equation}
	which combined with \cref{eq-kleg_high_derivative_in_kerL} yields:
	\begin{equation}
		c_r = 0, \quad \forall r\in\llbracket 0, k-1 \rrbracket.
	\end{equation}

	The end of the proof is now identical to that of \cref{lem_non_stability_klegs}.
	From $\psi_0(z) = P(z) \in \C_{k(k-2)-1}[z]$,
	we combine the relation $\psi_0 \perp Q(z^k) e^{\omega^r \alpha z}$
	for all $Q\in\C[X]$ and $r\in \llbracket 0, k-1 \rrbracket$
	with the fact that
	$ \braket{ \psi_0 \, | \, z^{k(k-3)} e^{\omega^r \alpha z}}
	= \braket{ \partial_z^{k(k-3)} \psi_0 \, | \, e^{\omega^r \alpha z}}
	= \overline{\partial_z^{k(k-3)} \psi_0( \overline{\omega^r \alpha})}$
	to deduce that $\partial_z^{k(k-3)} \psi_0$
	is a polynomial of degree at most $k-1$ and vanishing at the $k$ distinct points
	$\overline{\omega^r \alpha}$ for $0\leq r\leq k-1$;
	thus $\partial_z^{k(k-3)} \psi_0$ is identically null
	and $\psi_0$ is actually a polynomial of degree at most $k(k-3)-1$.
	Iterating this argument for all the derivatives
	$\partial_z^{k(k-j)}\psi_0$ with $3\leq j \leq k$,
	we finally obtain
	\begin{equation}
		\psi_0 = 0,
	\end{equation}
	which concludes the proof.

\clearpage
\section{Conclusion and perspectives}
\label{sec-conclusion}
We obtained a set of sufficient conditions to establish the convergence
of Lindblad master equations modeling bipartite open quantum systems used in reservoir
engineering approaches.
These conditions essentially ensure that energy stays bounded along trajectory
and that an algebraic question of density is satisfied, and we showed that they can be tested
on an explicit example by applying \cref{th-main_convergence}
to a Lindblad equation modeling multi-photon processes used for the stabilization
of cat qubits.
Even though the family of Lindblad equations we consider is physically motivated by
the use of adiabatic elimination to transfer the experimental difficulty from dissipation engineering
to Hamiltonian coupling engineering,
our analysis does not actually rely on this approximation,
and thus opens the way to the study of these systems outside the adiabatic regime.
\newline

In future work, several lines of research would be worth developing.

First, the general philosophy of reservoir engineering consists in engineering exotic
couplings to an ancillary dissipative buffer system,
with relative freedom on the exact nature of the buffer.
Intuitively, we expect natural conditions to be that the buffer system
converges to a unique steady-state when isolated,
and that the engineered interaction cancels on this state.
We thus intend to study the generalization of \cref{th-main_convergence} to the following
family of Lindblad equations:
\begin{equation}
	\label{eq-setting_generalise}
	\frac d{dt} \orho_t
		= -i \, [ \oL \oB^\dag + \oL^\dag \oB, \orho_t ]
			+ \mathcal L_B(\orho_t)
\end{equation}
where the Hilbert space has a tensor form $\H = \H_a \otimes \H_b$
with $\H_a$ and $\H_b$ two Hilbert spaces,
$\oL$ is an operator on $\H_a$,
$\mathcal L_B$ is an unspecified Lindblad generator acting on $\H_b$
such that the Lindblad equation $\frac d{dt} \orho^b_t = \mathcal L_B(\orho^b_t)$
has a unique and attractive steady state $\orho^b_\infty$,
and $\oB$ is an operator on $\H_b$ satisfying $\oB\orho^b_\infty = 0$.
In this setting, the question would be to determine sufficient conditions
ensuring that the solutions to \cref{eq-setting_generalise}
converge to density operators of the form
$\orho_\infty = \orho^a_\infty \otimes \orho^b_\infty$
with $\orho^a_\infty$ supported on $\ker\oL$.

Another natural (and possibly simple) extension would be to consider systems coupled
to multiple dissipative buffers,
used to approximate several dissipative processes on the target system.

Regarding the study of
engineered multi-photon processes,
a key fact is that the proof of the density criterion in $\H_a\otimes \H_b$
could be reduced to proving the density of another subspace in $\H_a$ only.
It is thus natural to wonder whether the density of this subspace is related to
the fact that the Lindblad equation
\begin{equation}
	\label{eq-ccl_onemode}
	\frac d{dt} \orho_t = D[\oL](\orho_t)
\end{equation}
converges to density operators supported on $\ker\oL$,
as was shown in \cite{azouitWellposednessConvergenceLindblad2016}
using Lyapunov arguments.
If one tries to formally replicate the reasoning in \cref{th-main_convergence}
for the study of \cref{eq-ccl_onemode},
the density condition states that the set
\begin{equation}
	\Span{ P(\oL^\dag \oL, \oL^\dag) \ket v
			\; \mid \;
			P\in\C\langle X,Y \rangle, \, \ket v\in\ker\oL}
\end{equation}
is dense in $\H_a$.
This is in fact closely related to the density result in $\H_a$ used in our proof,
where we
established a stronger result, using only polynomials $P(\oL^\dag \oL, \oL^\dag)$
of degree at most one in their first variable.
A natural question is to determine sufficient conditions
under which the convergence of \cref{eq-maineq} on $\H_a\otimes \H_b$
could be deduced directly from the convergence of \cref{eq-ccl_onemode} on $\H_a$,
fully justifying the use of adiabatic elimination.

Finally, the convergence result we obtain is qualitative,
as opposed to the quantitative convergence rates obtained with Lyapunov tools for instance
in \cite{azouitWellposednessConvergenceLindblad2016}.
Since the reservoir engineering setting systematically splits the system under study
into two subsystems, one of which only is dissipative,
one can wonder whether inspiration could be drawn from the study of
hypocoercive systems.
While such ideas are mentioned in \cref{sec-cat_compactness}
to clarify the intuition behind specific proof steps,
we failed so far to exploit these ideas to derive quantitative results
based for instance on a Lyapunov analysis.

\paragraph{Acknowledgements.}
The authors thank
Claude Le Bris,
Zaki Leghtas,
Mazyar Mirrahimi and
Alain Sarlette
for enlightening discussions and comments.

\section*{Declarations}
\paragraph{Data availability statement.}
This article has no associated data.
\paragraph{Funding and competing interests.}
This project has received funding from the European Research Council (ERC)
under the European Union’s Horizon 2020 research and innovation program
(grant agreement No. 884762).

\appendix

\clearpage
\printbibliography

\end{document}